\pdfoutput=1
\documentclass[10pt,twocolumn,journal]{IEEEtran}
\usepackage{latexsym}
\usepackage{float}
\usepackage{amsfonts}
\usepackage{amsbsy}
\usepackage{amssymb}
\usepackage{times}
\usepackage{graphicx}
\usepackage{enumerate}
\usepackage[usenames]{color}
\usepackage[dvips]{pstcol}
\usepackage{epstopdf}
\usepackage[caption=false]{subfig}
\usepackage{cite}
\usepackage{amssymb}
\usepackage{amsfonts}
\usepackage{graphicx}
\usepackage{epsfig}
\usepackage{psfrag}
\usepackage{xcolor}
\usepackage{amsfonts, bm}
\usepackage{epstopdf}
\usepackage{cite}
\usepackage{color}
\usepackage{xcolor}
\usepackage{subfig}
\usepackage{verbatim}
\usepackage{multirow}
\usepackage{booktabs}
\usepackage{amsthm}
\usepackage{makecell}
\usepackage{units}
\usepackage{algpseudocode}
\usepackage{amsmath}

\usepackage{stfloats}

\IEEEoverridecommandlockouts

\usepackage[linesnumbered,ruled]{algorithm2e}

\definecolor{ccr}{RGB}{0,0,255}  
\definecolor{ccb}{RGB}{0,0,255}  
\usepackage{hyperref}
\hypersetup{hypertex=true,
	colorlinks=true,
	linkcolor=ccr,
	anchorcolor=ccr,
	citecolor=ccb}

\begin{document}
	\title{Coarse-to-Fine: A Dual-Phase Channel-Adaptive Method for Wireless Image Transmission}
	\author{Hanlei Li, Guangyi Zhang, Kequan Zhou, Yunlong Cai, and Guanding Yu
		\thanks{Part of this work has been presented at the IEEE Wireless Communications and Networking Conference, Dubai, United Arab Emirates, April 2024 \cite{DRJSCC}.
		
		H. Li, G. Zhang, K. Zhou,  Y. Cai,  and G. Yu are with the College of Information Science and Electronic Engineering, Zhejiang University, Hangzhou 310027, China (e-mail: hanleili@zju.edu.cn; zhangguangyi@zju.edu.cn; kqzhou@zju.edu.cn; ylcai@zju.edu.cn; yuguanding@zju.edu.cn).	
}}
	
	\maketitle
	
	\begin{abstract}
Developing channel-adaptive deep joint source-channel coding (JSCC) systems is a critical challenge in wireless image transmission. While recent advancements have been made, most existing approaches are designed for static channel environments, limiting their ability to  capture the dynamics of channel environments. As a result, their performance may degrade significantly in practical systems. In this paper, we  consider time-varying block fading channels, where the transmission of a single image can experience multiple fading events. We propose a novel coarse-to-fine channel-adaptive JSCC framework (CFA-JSCC) that is designed to  handle both significant fluctuations and rapid changes in wireless channels. Specifically, in the coarse-grained phase, CFA-JSCC utilizes the average signal-to-noise ratio (SNR) to adjust the encoding strategy, providing a preliminary adaptation to the prevailing channel conditions. Subsequently, in the fine-grained phase, CFA-JSCC  leverages instantaneous SNR to dynamically refine the encoding strategy. This refinement is achieved by re-encoding the remaining channel symbols whenever the channel conditions change.  Additionally, to reduce the overhead for SNR feedback, we
utilize a limited set of channel quality indicators (CQIs) to represent the channel SNR and further propose a reinforcement learning (RL)-based CQI selection strategy to learn this mapping.
This strategy incorporates a novel reward shaping scheme that provides  intermediate rewards to facilitate the training process. Experimental results demonstrate that our CFA-JSCC provides enhanced flexibility  in capturing channel variations and improved robustness  in time-varying channel environments.
	\end{abstract}
	

	\section{Introduction}
	
	The deployment of substantial intelligent devices, such as Internet of Things (IoT) sensors and autonomous vehicles \cite{iot,data}, has led to a rapid surge in the need to transmit vast amounts of data. To address this challenge, there is an increasing  demand  for communication systems  with  enhanced intelligence that  can adaptively recognize and process information. This demand has progressively stimulated research interest in integrating deep learning (DL) into the design of wireless transceivers, particularly in the development of semantic communication systems based on deep joint source-channel coding (JSCC) \cite{TWC_1,TWC_2,TWC_3}.
	\subsection{Prior Work}
	The main advances in the DL-based JSCC  systems have been driven by the success of  autoencoders  parameterized by  deep neural networks (DNNs) \cite{DeepJSCC,Deep_F,perc,BI,domain,JSCC_text,Deep_video,Deep_mul,JSCC_speech,JSCC_perc,TASK_O,DL_perc,HARQ}. Unlike traditional separation-based schemes, this approach employs DNNs to jointly map input source data  to channel symbols,  eliminating the need for  explicit source and channel codecs, as well as modulation operations. Inspired by this concept, DeepJSCC \cite{DeepJSCC} has emerged as one of the technical foundations for wireless image transmission. It has been verified to outperform classical separation-based methods, demonstrating  significantly better robustness against cliff effects. 
	Furthermore, the authors of \cite{Deep_F} have integrated channel output feedback into the JSCC system to enhance image reconstruction
	quality effectively.
    Additionally, \cite{perc} has introduced a novel JSCC system that accounts for human visual perception. This approach integrates perceptual loss and adversarial loss to capture both global semantic information and local texture, thereby improving the perceptual quality of reconstructed images.
    In \cite{BI}, an adaptive information bottleneck (IB)-guided JSCC method has been proposed to reduce transmission overhead and enhance image reconstruction quality simultaneously.  Moreover, a domain knowledge-driven semantic communication system  has been proposed in \cite{domain}, featuring a dual-path framework that considers both semantic extraction and reconstruction at the information and concept levels.

	Although the aforementioned works  have demonstrated success, they primarily focus on optimizing models for fixed signal-to-noise ratio (SNR) levels. Consequently, their performance often degrades when SNR mismatches occur between the training and deployment phases. To address this  limitation, several solutions have been  proposed \cite{ADJSCC,DeepJSCC_V,SCAN,OFDM,CAJSCC,MITT,DL_MMSE,
	DeepJSCC_l,ADJSCC_l,swin_taming}. Specifically, the authors of \cite{ADJSCC} have introduced an attention DL-based JSCC (ADJSCC) model that incorporates SNR as an input to guide the encoding process. This approach enables effective operation across varying SNR levels during transmission, offering superior robustness against channel variations. Building upon \cite{ADJSCC}, 
	\cite{DeepJSCC_V} has proposed a predictive and adaptive deep coding framework that jointly considers channel SNR, coding rate, and image content to achieve adaptive coding rates tailored to different channel conditions.
	Furthermore, other approaches have explored more practical system settings. For instance,
	\cite{SCAN} has presented an adaptive image transmission system  designed for multiple-input multiple-output (MIMO) channels. This framework utilizes the channel state information (CSI) matrix and noise variance to generate an attention mask, prioritizing  critical features by allocating them higher transmission power. In \cite{OFDM}, the authors have introduced a pioneering JSCC scheme that integrates an autoencoder with orthogonal frequency division multiplexing (OFDM) and employs CSI for adaptive power allocation. 
	Expanding on this concept, \cite{CAJSCC} has proposed a channel-adaptive JSCC (CAJSCC) model for OFDM transmission, incorporating a novel dual-attention mechanism to leverage estimated CSI for optimized power resource allocation.
	\begin{figure}[t]
		\centering 
		\includegraphics[width=1.0\linewidth]{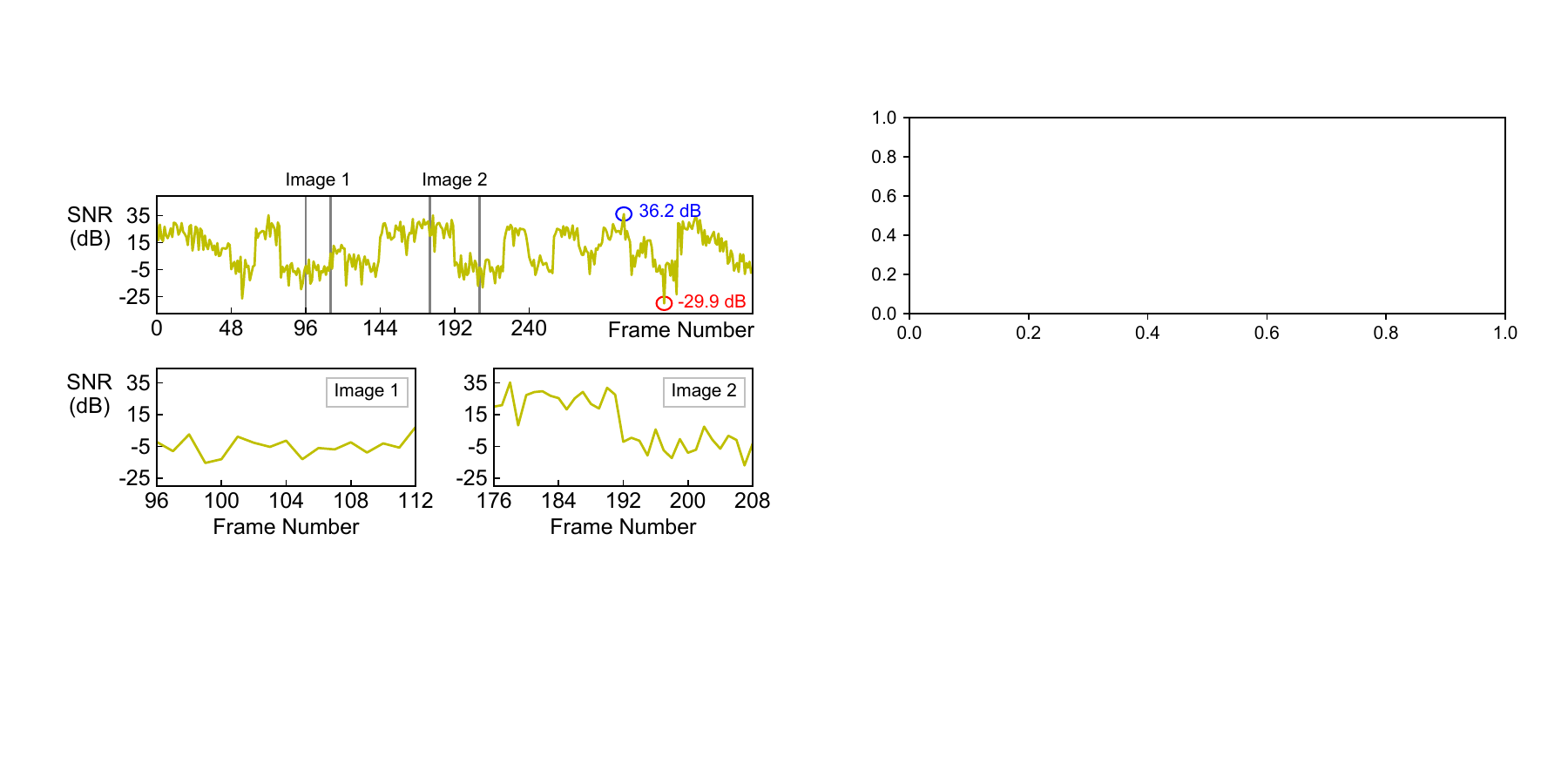} 
		\captionsetup{font={footnotesize  }}
		\captionsetup{justification=raggedright,singlelinecheck=false}
		\caption{
			Illustration of the time-varying communication scenarios. The top subfigure shows the SNR versus frame number. The two subfigures below highlight the transmissions of image 1 and  image 2, respectively.
			}
		\label{fig:snr} 
	\end{figure}
	\subsection{Motivation and Contributions}

	Despite the progress made by existing methods, two key limitations can be identified:
	\begin{itemize}
			\item[(i)]
		\textit{Inability to  manage  significant channel fluctuations:} In real communication systems, image transmission  is often subject to  significant fading events, as shown in Fig. \ref{fig:snr}. However, most existing models rely on a single attention module to capture channel variations. These methods lack the capacity to handle a wide range of channel SNRs simultaneously, making them less effective at dealing with significant channel fluctuations.
		Furthermore, when trained under conditions of significant channel fluctuations,  reliance on a single attention module can lead to unstable gradient updates, impeding model convergence and undermining training stability. 
	
	\item[(ii)]
	\textit{Limitation to support varying channel conditions:} In real-world scenarios, both  image code lengths
	and channel coherence time are uncertain.
	Channel conditions can vary even during the transmission of a single image, as  illustrated in Fig. \ref{fig:snr}. 
	However, existing models, such as ADJSCC, typically rely on a fixed SNR  to optimize  encoding strategies for channel adaptation. Given that a 
	fixed SNR only reflects the channel condition as a statistical average and  fails to account for real-time variations, the existing models tend to fail in practical
	scenarios  characterized by rapidly varying channels.
		\end{itemize}

In this paper, we investigate  channel-adaptive system  designs that simultaneously cope with significant  fluctuations and rapid changes  in wireless channels.  Unlike previous works, we consider time-varying block fading channels, where the transmission of a single image may undergo multiple fading events. Subsequently, similar to \cite{HJSCC}, we extend the hierarchical variational autoencoder (VAE) architecture \cite{HVAE} to the OFDM system as our backbone model,  leveraging hierarchical VAE’s strong  ability to incorporate contextual information for managing channel fluctuations. To address the two limitations, starting from the backbone, we propose a novel coarse-to-fine channel-adaptive JSCC framework (CFA-JSCC), which divides the encoding process into two phases: the coarse-grained phase and the fine-grained phase. In particular, the coarse-grained phase aims to  handle  significant channel fluctuations, while
the fine-grained phase targets the rapid changes in wireless channels.

To implement the proposed coarse-to-fine mechanism, we design a dual-phase channel-adaptive strategy. In the coarse-grained phase, the average SNR is introduced to assess long-term channel conditions. CFA-JSCC  incorporates an attention module  that adjusts the learned features based on the average SNR, 
enabling preliminary adaptation to the channel environment.  Subsequently, in the fine-grained phase, the instantaneous SNR is  introduced to capture real-time channel fluctuations. Whenever the channel conditions change, CFA-JSCC leverages the instantaneous SNR to re-encode the channel symbols, thereby refining the encoded features to  better align with the current channel state. By utilizing the dual-phase strategy, CFA-JSCC exhibits improved robustness against time-varying channels. 

To reduce the overhead for SNR feedback, we  explore the challenge of efficiently mapping SNR to finite channel quality indicators (CQIs). To address this, we model the block fading channel as a Markov chain with unknown transition probabilities. Then, we formulate the CQI selection strategy as a Markov Decision Process (MDP) and develop a reinforcement learning (RL)-based method to address the problem. Since immediate rewards are unavailable for RL in this context,  we propose a novel reward shaping scheme that provides  intermediate rewards to facilitate the training process. Leveraging this method, we significantly reduce the feedback overhead while  minimizing performance loss. Simulation results  demonstrate that the proposed scheme  enhances flexibility  in capturing channel variations and significantly  improves overall performance  under varying channel conditions.

	The main contributions of this paper can be summarized as follows:
	\begin{itemize}
\item We  explore the channel-adaptive  JSCC design in   block fading channels, where the transmission of a single image can experience multiple fading events. To seamlessly adapt to time-varying channels,  a  coarse-to-fine channel-adaptive JSCC model, namely CFA-JSCC, is proposed.

\item We design a dual-phase channel-adaptive strategy to facilitate the  implementation of  CFA-JSCC. The system first adjusts the learned image features in a coarse-grained manner based on the average SNR, followed by a fine-grained optimization of the encoding strategy using the instantaneous SNR.

\item We design an RL-based CQI selection strategy to reduce the overhead  for SNR feedback. This strategy incorporates a novel reward shaping scheme that provides  intermediate rewards,  facilitating the training process.

\item  We  present extensive experimental results to verify the effectiveness of our method.  Compared to existing DL-based JSCC models, the proposed scheme  demonstrates enhanced flexibility  in capturing channel variations and improved robustness against time-varying channels.

	\end{itemize}

\subsection{Organization and Notations}
The rest of this paper is organized as follows. Section II introduces the framework of CFA-JSCC. Section III describes  the detailed architecture of the proposed channel-adaptive system. Section IV provides a comprehensive description of the RL-based CQI selection strategy. Simulation results are presented in Section V. Finally, Section VI concludes the paper.

$\textit{Notations:}$ Scalars, vectors, and matrices are respectively denoted by lower case, boldface lower case, and boldface upper case letters. For a scalar $a$, $|a|$
 represents its absolute value. For a vector $\bm{a}$, $||\bm{a}||$ is its Euclidean norm. Finally, 
${\mathbb{C}^{m \times n}}\;({\mathbb{R}^{m \times n}})$ are the space of ${m \times n}$ complex (real) matrices.

	\section{Coarse-to-Fine Channel-Adaptive Framework}
	In this section,  we begin with a system overview of the CFA-JSCC. Next, we present the specific channel transmission process employing  OFDM. Subsequently, we detail the implementation of the coarse-to-fine channel-adaptive coding.
    \subsection{System Overview}

    We commence by providing an overview of our proposed CFA-JSCC, which is presented in Fig. \ref{fig:typical_JSCC}.
    In particular, the transmitter consists of a bottom-up path and a top-down path to create multiple hierarchical representations of the input image, while the receiver utilizes a top-down path for image reconstruction. The simplified JSCC encoders/decoders without the  coarse-to-fine mechanism are employed in this subsection to enhance clarity.
    
    Given an input image $\bm{x}\in \mathbb{R}^{N}$, the bottom-up path first utilizes $\bm{x}$ to generate a set of latent features, where $N$ denotes the number of pixels. These latent features are then processed layer-by-layer through the latent module along the top-down path, autoregressively generating latent representations  $\bm{\mu} =\left\{\bm{\mu}_1, \bm{\mu}_2, \ldots, \bm{\mu}_L\right\}$.  Subsequently, these representations are fed to a set of JSCC encoders. In this way, we obtain a set of channel symbol vectors $\bm{s} =\left\{\bm{s}_1, \bm{s}_2, \ldots, \bm{s}_L\right\}$, where $\bm{s}_l \in \mathbb{C}^{K_l}$ for $l=1,2,\ldots,L$ denotes the channel symbol vector of the $l$-th layer, and $K=\sum_{l=1}^{L}K_l$  represents the total  number of channel symbols. Accounting for the limited
    transmission power, the channel symbol vector of each layer is subject to a power constraint $P$, i.e., $\frac{1}{K_l}  \mathbb{E} ||\bm{s}_l||^2 \leq P$ for $l= 1,2,\ldots,L$. Then, these channel symbol vectors are  transmitted to the receiver through   wireless channels.
    
    At the receiver, the  noisy channel symbol vectors undergo processing by a set of JSCC decoders, and we obtain the recovered representations $\bm{\hat{\mu}} =\left\{\bm{\hat{\mu}}_1, \bm{\hat{\mu}}_2, \ldots, \bm{\hat{\mu}}_L\right\}$. With $\bm{\hat{\mu}}$ at hand, a top-down path in the receiver is employed to reconstruct image $\bm{\hat{x}}_H$. The distortion between the original image $\bm{x}$ and the reconstructed image $\bm{\hat{x}}_H$ is defined using mean square-error (MSE), which is given by
    \begin{equation} \label{loss}
    	d(\bm{x},\bm{\hat{x}}_H) = \dfrac{1}{N}||\bm{x}-\bm{\hat{x}}_H||^{2}.
    \end{equation}
    Additionally, we introduce the channel bandwidth ratio (CBR) to describe the transmission rate (overhead), which is expressed as $\text{CBR}=K/N$. Under a certain CBR, the goal is to determine the encoder and decoder parameters that minimize the expected distortion defined in Eq.  (\ref{loss}).
    
   \subsection{Transmission with OFDM } 
    
\begin{figure}[t]
     	\centering 
     	\includegraphics[width=1.0\linewidth]{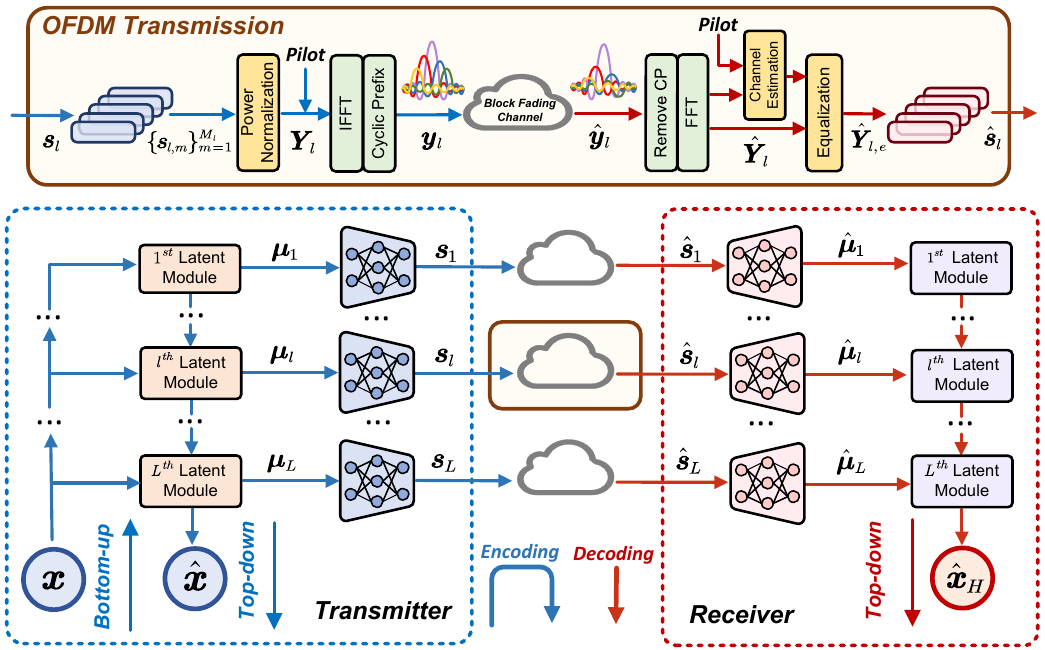} 
     	\captionsetup{font={footnotesize  }}
     	\captionsetup{justification=raggedright,singlelinecheck=false}
     	\caption{The system overview of CFA-JSCC. The transmitter includes a bottom-up path and a top-down path, while the receiver includes a top-down path.}
     	\label{fig:typical_JSCC} 
     \end{figure}

      Unlike prior studies, we consider  time-varying block fading channels, where the transmission of a single image  may  experience multiple fading events. Therefore, in this subsection, we provide a detailed  description   of the channel transmission process  in CFA-JSCC,   as illustrated in the top part of Fig. \ref{fig:typical_JSCC}. For clarity, we focus on  the $l$-th layer of the model for further  explanation, as the operation of other layers follows a similar process.
     
      We consider an OFDM-based JSCC system operating over
      the block fading channels. We first divide the output of the JSCC encoder, $\bm{s}_l$,   into $M_l$ blocks to adapt to the block fading channels. This process  is described as $\bm{s}_l =\{{\bm{s}_{l,m}}\}^{M_l}_{m=1}$,
     where $\bm{s}_{l,m}$ represents the $m$-th block.
     Subsequently, each block is transmitted using $N_s$ OFDM symbols, $N_p$ pilot symbols, and $L_f$ subcarriers.
     Taking block $\bm{s}_{l,m}$ as an example, we first reshape it into the complex matrix $\bm{Y}_{l,m} \in \mathbb{C}^{N_s\times L_f}$.
      Then, we apply the inverse fast Fourier transform (IFFT) to each OFDM symbol and append a cyclic prefix (CP) of length $L_{cp}$.
        The resulting time-domain symbols $\bm{y}_{l,m} \in \mathbb{C}^{(N_s + N_p)(L_f + L_{cp})}$ are transmitted through wireless channels.
       This process  is represented as $\bm{\hat{y}}_{l,m}=\bm{h}_{l,m}*\bm{y}_{l,m}+\bm{w}_{l,m}$, where $\bm{\hat{y}}_{l,m}$ denotes the received time-domain symbols  for the $m$-th block, $*$ represents the linear convolution operation,
        $\bm{h}_{l,m} \in \mathbb{C}^{L_t}$ is the channel impulse response with $L_t$ being the number of multipaths, and $\bm{w}_{l,m} \sim \mathcal{CN}(0,\sigma^2\bm{I})$  denotes additive white Gaussian noise (AWGN).
     
        \begin{figure*}[t]
     	\centering 
     	\includegraphics[width=0.83\linewidth]{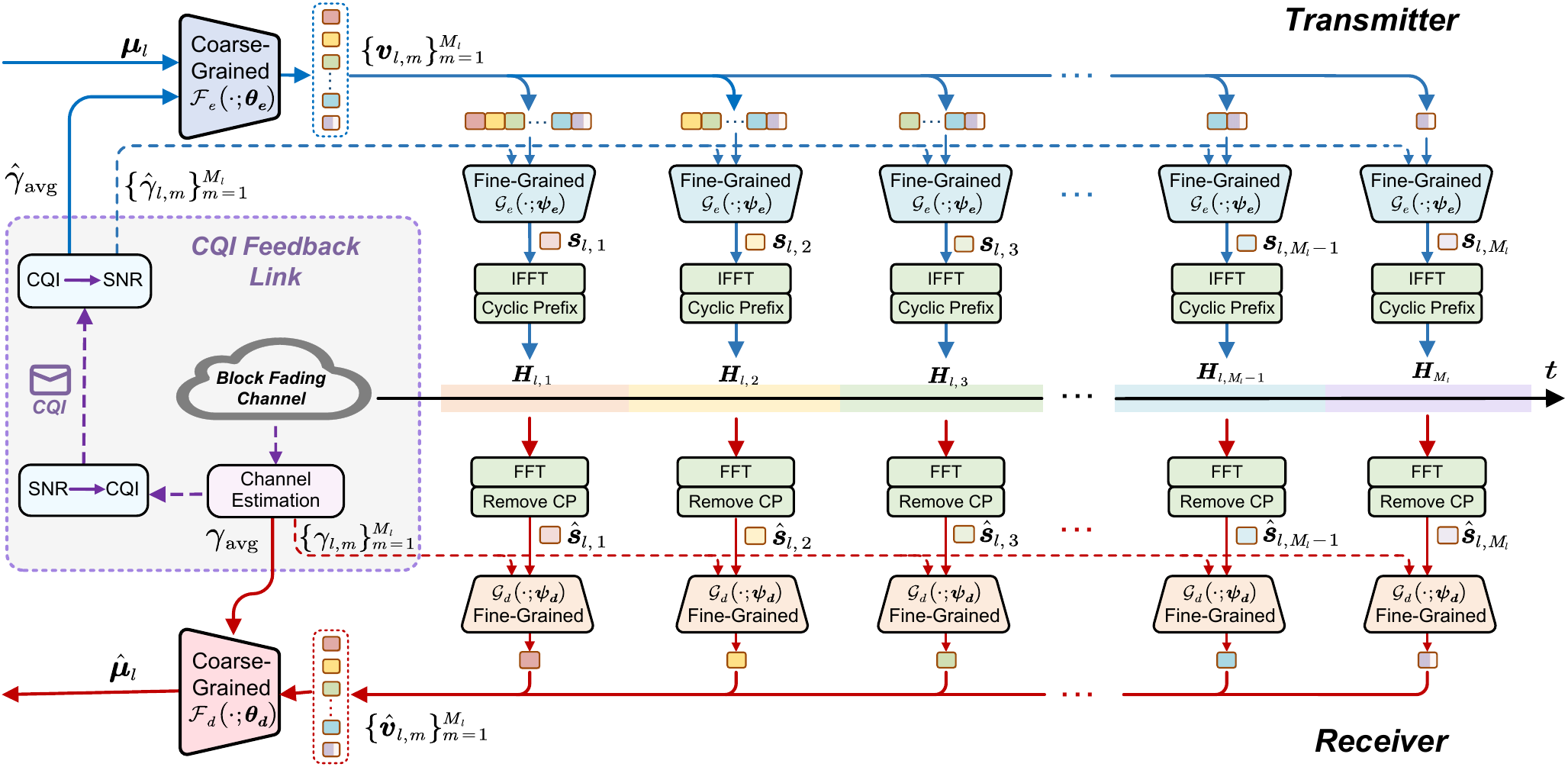} 
     	\captionsetup{font={footnotesize  }}
     	\captionsetup{justification=raggedright,singlelinecheck=false}
     	\caption{The process of coarse-to-fine channel-adaptive coding.}
     	\label{fig:system_model} 
     \end{figure*}
     
     	At the receiver, upon  obtaining the noisy channel output $\hat{\bm{y}}_{l,m}$,  the CP  is removed, and the fast Fourier transform (FFT) is applied to each OFDM symbol to generate the frequency-domain symbols $\bm{\hat{Y}}_{l,m}$. The relationship between the transmitted and received  symbols at the $k$-th subcarrier of the $t$-th OFDM symbol within  the $m$-th block in the frequency domain is expressed as
     	\begin{equation} \label{OFDM}
     		\bm{\hat{Y}}_{l,m}[t,k]=\bm{H}_{l,m}[k,k]  \bm{Y}_{l,m}[t,k] + \bm{W}_{l,m}[t,k],
	    \end{equation}
     	where $\bm{H}_{l,m} \in \mathbb{C}^{L_f \times L_f}$ is a diagonal matrix representing  the channel frequency response, and $\bm{W}_{l,m} \in \mathbb{C}^{N_s\times L_f} $ is the frequency-domain noise matrix. 
     	
     	Next, utilizing the received pilot symbols, we employ the linear minimum mean square-error  (LMMSE)  channel estimation technique to estimate $\bm{H}_{l,m}$. Following this, a zero-forcing (ZF) equalizer is  applied to individually equalize each block based on the estimated $\bm{H}_{l,m}$. Upon aggregating all blocks, we obtain  the noisy channel symbol vector $\bm{\hat{s}}_l$, which is then processed by the JSCC decoder along with the top-down path to obtain the reconstructed image.

        To characterize the channel conditions  in block fading channels, we introduce both average SNR and instantaneous SNR. The average SNR, $\gamma_{\textrm{avg}}$, is  defined as
        \begin{equation} \label{avg_snr}
        	\gamma_{\textrm{avg}}=10 \log \frac{P}{\sigma^2}(\text{dB}).
        \end{equation}
         For the instantaneous SNR, $\gamma_{l,m}$, of the $m$-th block at the $l$-th layer, 
        we account for the varying SNR of each subcarrier in the OFDM system by employing the effective exponential SNR mapping (EESM) algorithm \cite{EESM}. This algorithm consolidates the subcarrier-specific SNRs into a single effective SNR value, expressed as
        \begin{equation} \label{inst_snr}
        	\gamma_{l,m}=-\beta \log \left(\frac{1}{L_f}\sum_{k=1}^{L_f} e^{-\frac{\gamma^k_{l,m}}{\beta}}\right) (\text{dB}),
        \end{equation}
        where $\beta$ is an adjustable parameter, which is set to $5$, $\gamma^k_{l,m}= 20 \log ( {h^k_{l,m}}\sqrt{P}/{\sigma})(\text{dB})$ represents the SNR of the  $k$-th subcarrier, and $h^k_{l,m}$ denotes the  channel gain coefficient  of the  $k$-th subcarrier.

     \subsection{Coarse-to-Fine Channel-Adaptive Coding } 	
     In this paper, we concentrate on designing a channel-adaptive JSCC model capable of addressing both significant fluctuations and rapid changes in wireless channels. To this end, we further design the coarse-grained encoder/decoder and fine-grained encoder/decoder, as illustrated in Fig. \ref{fig:system_model}. We focus on the $l$-th layer
     of the model  for clarity.

     Given the output of the top-down path $\bm{\mu}_l$,
     the coarse-grained encoder first encodes $\bm{\mu}_l$ and  ${\hat{\gamma}_{\textrm{avg}}}$ to generate a set of   symbols $\bm{v}_l$, where ${\hat{\gamma}_{\textrm{avg}}}$ is the  received
     average SNR  from the feedback link.
  	 These symbols are then divided into $M_l$ blocks to adapt to the block fading channels. This process can be expressed as
     	\begin{equation} \label{coarse_encode}
     		\bm{v}_l \triangleq\{{\bm{v}_{l,m}\}^{M_l}_{m=1}} = \mathcal{F}_{e}(\bm{\mu}_l,\hat{\gamma}_{\textrm{avg}};\bm{\theta}_e),
     	\end{equation}
     	where $\mathcal{F}_e$  represents the coarse-grained encoding function, and  $\bm{\theta}_e$ are its trainable parameters.
     	
     	Subsequently, due to rapid variations in the fading channel, capturing channel characteristics effectively using only the coarse-grained encoding becomes challenging.
     	To address this, we propose a fine-grained encoder that utilizes  $\hat{\gamma}_{l,m}$  to dynamically optimize the encoding strategy to adapt to the varying channel conditions. Here, $\hat{\gamma}_{l,m}$ represents the received instantaneous SNR of the $m$-th block obtained from the feedback link.
     	Specifically, the set of frequency-domain channel gain responses is denoted as $\left\{\bm{H}_{l,1}, \bm{H}_{l,2}, \ldots, \bm{H}_{l,M_l}\right\}$, where $\bm{H}_{l,m}=\textrm{diag} \left(h^1_{l,m}, h^2_{l,m}, \ldots, h^{L_f}_{l,m}\right)$ for $m=1,2,\ldots,M_l$ represents  the channel gain response of the $m$-th block. 
     	As channel conditions change, the fine-grained encoder uses the remaining untransmitted blocks and the instantaneous SNR to generate a new block that adapts to the current channel state, as shown in Fig. \ref{fig:system_model}.
     	For example,  if  the channel gain response shifts from  $\bm{H}_{l,1}$ to $\bm{H}_{l,2}$,   the fine-grained encoder takes the remaining untransmitted blocks $\left\{\bm{v}_{l,2},\bm{v}_{l,3},  \ldots, \bm{v}_{l,M_l}\right\}$ and $\hat{\gamma}_{l,2}$ as input, 
     	 producing the block ${\bm{s}_{l,2}}$, which is optimized for the new channel response $\bm{H}_{l,2}$.
     	 In this way, the set of blocks  $\bm{s}_l=\left\{\bm{s}_{l,1},\bm{s}_{l,2},  \ldots, \bm{s}_{l,M_l}\right\}$  is generated, with each  block optimized for its respective channel environment. These blocks are then transmitted to the receiver using OFDM, as described in the previous section. Each fine-grained encoding process can be summarized as
     	 \begin{equation}\label{fine_encoder}
     	 \bm{s}_{l,m}= \mathcal{G}_{e}(\{{\bm{v}_{l,i}\}^{M_l}_{i=m}},\hat{\gamma}_{l,m};\bm{\psi}_e),
     	\end{equation}
     	where $\mathcal{G}_e$  represents the fine-grained encoding function,  $\bm{\psi}_e$ are its trainable parameters, and $m$ to $M_l$ denote the sequence numbers of  the remaining untransmitted blocks, with $m= 1,2,\ldots,M_l$.

     	At the receiver, the received blocks undergo fine-grained decoding first. Similar to the encoding process, each fine-grained decoding step can be expressed as
     	\begin{equation} \label{fine_decoder}
     		\bm{\hat{v}}_{l,m}= \mathcal{G}_{d}(	\bm{\hat{s}}_{l,m},\gamma_{l,m};\bm{\psi}_d),
     	\end{equation}
     	where $\mathcal{G}_d$ represents the fine-grained decoding function, and $\bm{\psi}_d$ are its trainable parameters. 
     	Furthermore, we integrate all blocks and perform coarse-grained decoding based on  $\gamma_{\textrm{avg}}$, which can be described as
     	\begin{equation}  \label{coarse_decoder}
     		\bm{\hat{\mu}}_l  = \mathcal{F}_{d}(\{{\bm{\hat{v}}_{l,m}\}^{M_l}_{m=1}},\gamma_{\textrm{avg}};\bm{\theta}_d),
     	\end{equation}
     	where $\mathcal{F}_d$ denotes the coarse-grained decoding function, and $\bm{\theta}_d$ are its  trainable parameters.
     	
\begin{figure*}[t]
	\centering 
	\includegraphics[width=0.95\linewidth]{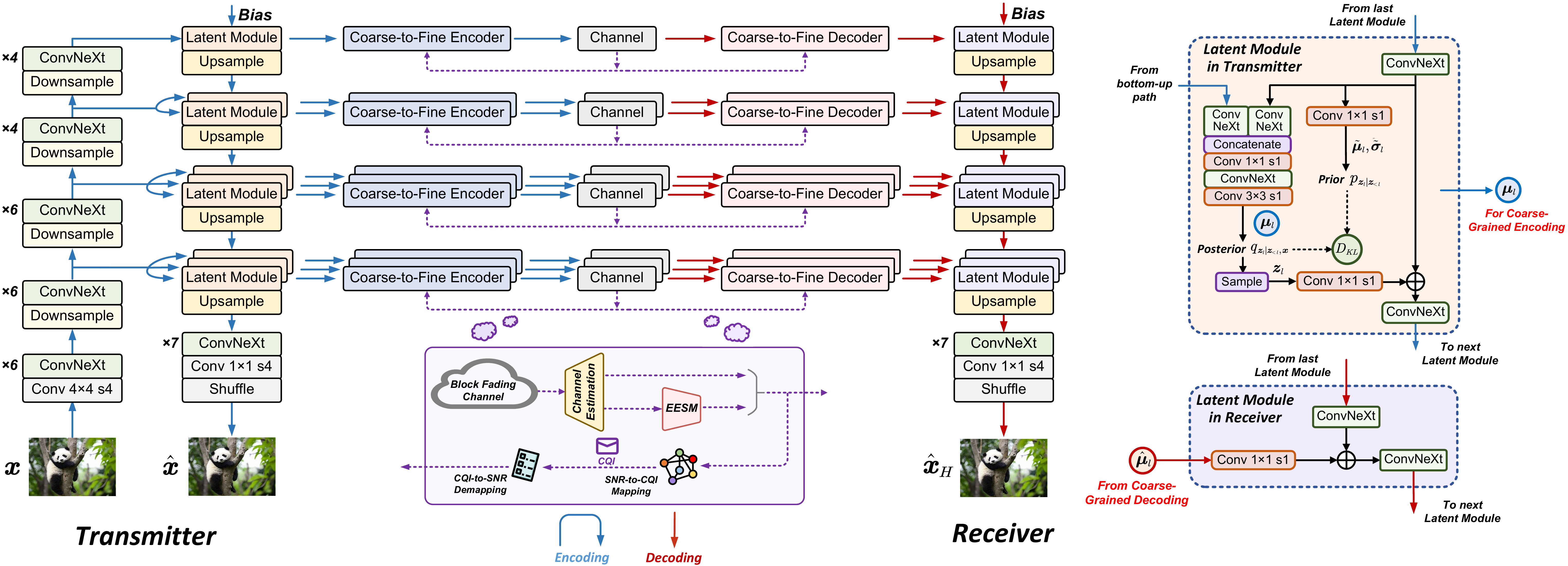} 
	\captionsetup{font={footnotesize  }}
	\captionsetup{justification=raggedright,singlelinecheck=false}
	\caption{The detailed architecture of  CFA-JSCC. At the transmitter, the model employs a bottom-up path followed by a top-down path to generate latent representations. Coarse-grained and fine-grained encoders are then utilized for  channel adaptation. The receiver adopts a similar structure to ensure seamless processing.  
	}
	\label{fig:Overall_architecture} 
\end{figure*}

    \section{Dual-Phase Channel-Adaptive Architecture}

	In this section, we provide a comprehensive introduction to the network architecture. Specifically, we first present the overall architecture of the proposed CFA-JSCC. Furthermore, we delve into the design specifics of the coarse-grained and fine-grained encoders/decoders. Finally, we detail the training algorithm employed for the CFA-JSCC.

	\subsection{Overall Architecture}
	Fig. \ref{fig:Overall_architecture} presents the overall architecture of CFA-JSCC. The model begins with a bottom-up path, where the input image is processed through multiple ConvNeXt modules \cite{conv_net} and downsample  layers, progressively generating latent features. This hierarchical  feature  extraction establishes  strong connections across layers, enabling the model to effectively leverage contextual information to adapt to varying channel conditions. Following this, a top-down path  further  processes  these features to generate latent representations, assisted by multiple latent modules and upsample layers. The upper right of Fig. \ref{fig:Overall_architecture} illustrates the specific architecture of the latent module in the encoder, which is  primarily based on ConvNeXt due to its superior performance in vision-related tasks. Subsequently, the output representations  from the top-down path  are  passed  through  coarse-grained  and fine-grained encoders to generate channel symbol vectors, which are then transmitted to the receiver  over wireless channels.
	
	At the receiver, decoding begins with  the fine-grained and  coarse-grained decoders, similar to the encoding process at the transmitter. The output representations from the coarse-grained decoders are then fed into the receiver’s top-down path, which shares identical parameters with the transmitter’s top-down path. Within  this path, latent representations are processed layer-by-layer, culminating in the reconstruction of the image.

       \begin{figure}[t]
      	\centering 
      	\includegraphics[width=0.99\linewidth]{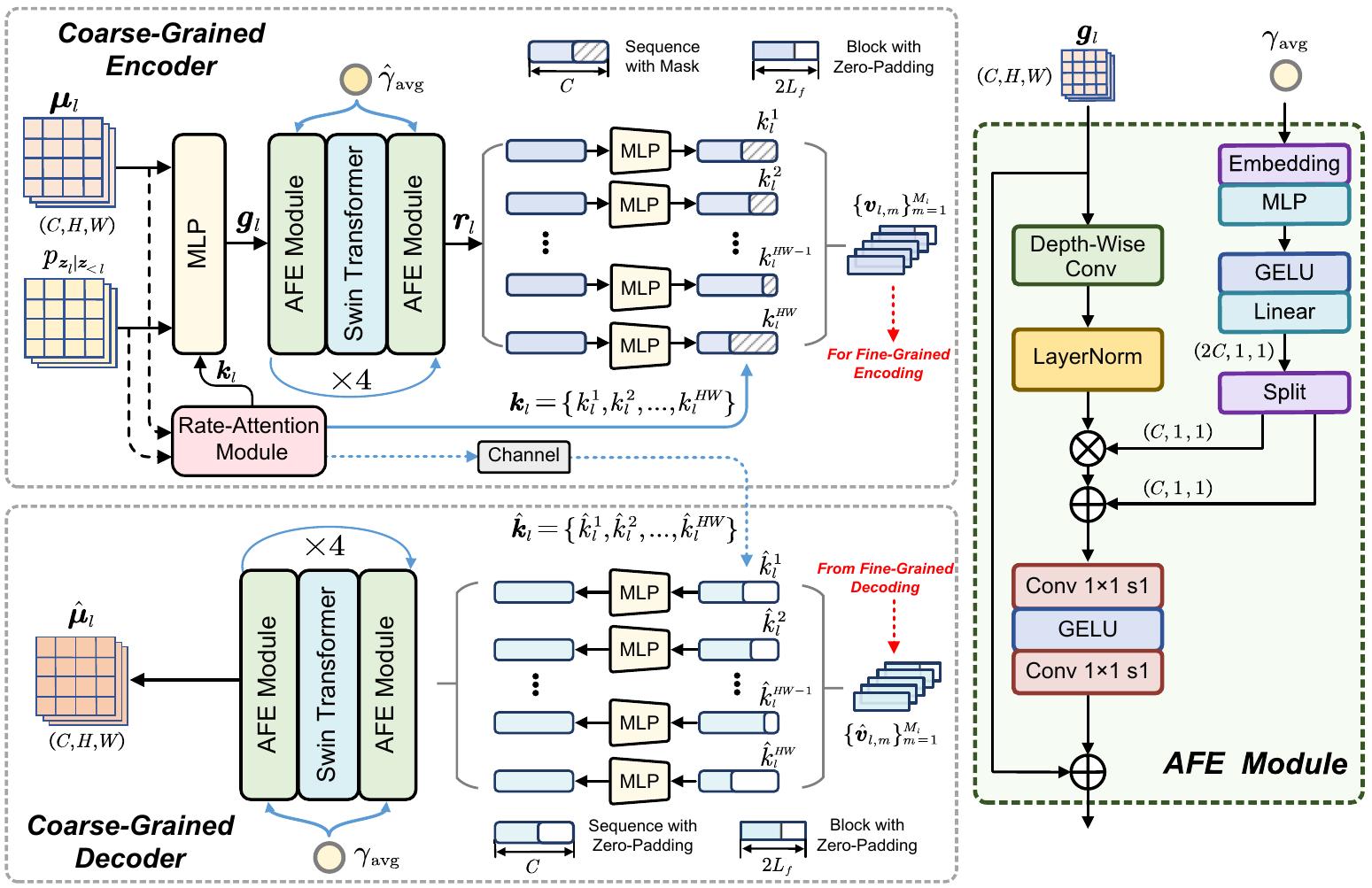} 
      	\captionsetup{font={footnotesize  }}
      	\captionsetup{justification=raggedright,singlelinecheck=false}
      	\caption{The architecture of the coarse-grained encoder and  decoder.}
      	\label{fig:Coarse_Encoder} 
      \end{figure}
         	  		\begin{figure*}[h]
         	\centering 
         	\includegraphics[width=0.94\linewidth]{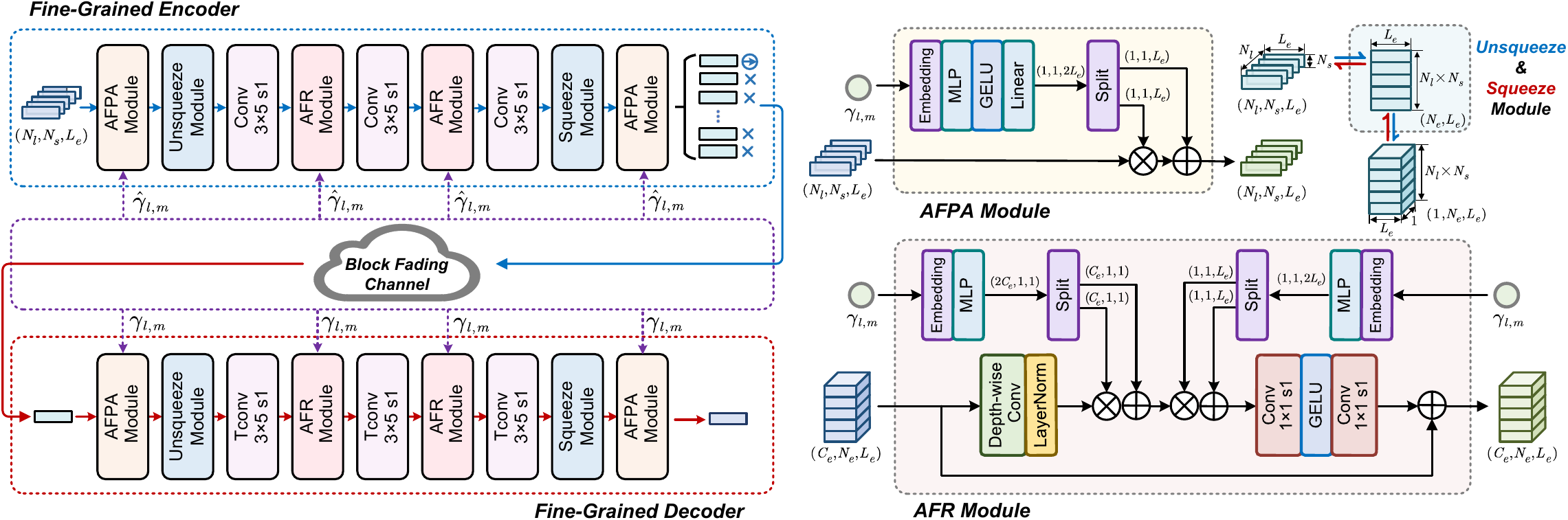} 
         	\captionsetup{font={footnotesize  }}
         	\captionsetup{justification=raggedright,singlelinecheck=false}
         	\caption{The architecture of the fine-grained encoder and  decoder.}
         	\label{fig:Fine_Encoder} 
         \end{figure*}
         
    \subsection{Coarse-Grained Coding Design}
    The coarse-grained encoder and decoder  leverage the average SNR  for coarse-grained coding, thereby achieving preliminary adaptation to  time-varying channels. The specific architecture is illustrated in Fig. \ref{fig:Coarse_Encoder}. Our focus is primarily on the design of the  coarse-grained encoder,  while  the decoder is  structured  as the  inverse of the encoder.

    \textit{1) Rate-Attention Module:} 
	The coarse-grained encoder begins with a rate-attention module for variable-length transmission.  It splits the input features into $HW$ sequences, each with a length of $C$, and then generates a set of permissible transmission lengths $\bm{k}_l=\{{k_{l}^o\}^{HW}_{o=1}}$ for all  sequences. Inspired by \cite{NTSCC}, the expression of $k_{l}^o$ is given by   $k_l^o=\sum_{c=1}^{C}- \alpha \log p_{\bm{z}_l^{(o,c)} | \bm{z}_{<l}}(\bm{z}_l^{(o,c)} | \bm{z}_{<l})$, where $\alpha$ is the adjustment coefficient and  $k_l^o$  actually represents the summation of entropy  across  all  $C$ dimensions of $\bm{z}_l^o$.  In our design,  $\bm{k}_l$ serves three critical functions. Firstly, it is utilized to generate a mask for the sequence,  facilitating variable-length transmission. Secondly,  $\bm{k}_l$ is  transmitted to the receiver in an error-free manner to aid in decoding. Thirdly,
	  $\bm{k}_l$  is fed into a multilayer perceptron (MLP) layer  along with the image features $\bm{\mu}_l$
	and the prior $p_{\bm{z}_l | \bm{z}_{<l}}$
	to perform preliminary feature processing. This process  is described as  $\bm{g}_l=\textit{MLP}(\bm{k}_l,\bm{\mu}_l,p_{\bm{z}_l | \bm{z}_{<l}})$, where $\textit{MLP}(\cdot)$ denotes the MLP layer.
	
	\textit{2) Coarse-Grained Channel Adaptation:} 
    With  the processed  features $\bm{g}_l$ at hand, we further design an attention feature embedding (AFE) module to perform coarse-grained adjustments based on the average SNR.
    Its architecture is illustrated in the right part of Fig. \ref{fig:Coarse_Encoder}. 
    Initially,  the depth-wise convolution with layer normalization is employed to preprocess  $\bm{g}_l$, which balances computational complexity and performance. Simultaneously, a sinusoidal embedding operation  converts the average SNR $\gamma_{\textrm{avg}}$ into a latent representation, which is further  refined by the MLP layer  to generate scaled  and shifted vectors. Using the layer-normalized features along with the generated scaled and shifted vectors,
    the model performs scaling and translation on the features, embedding the average SNR into the model. Then, the output is processed  through  convolutional layers and  a simple residual network \cite{ResNet} is incorporated  to accelerate convergence.   Beyond the AFE module, we also  integrate  the Swin Transformer blocks \cite{swin}, which we find particularly effective in obtaining robust channel symbols against noise. The input features $\bm{g}_l$  are sequentially processed through an alternating arrangement of AFE modules and Swin Transformer blocks, enabling preliminary adaptation to the channels.
    
    \textit{3) Masking and Regrouping:} 
     With the output features of the final AFE module  denoted as $\bm{r}_l=\{{\bm{r}_{l}^o\}^{HW}_{o=1}}$ and the set of permissible transmission lengths  $\bm{k}_l$, we utilize the MLP layer to process the features and mask the output according to $\bm{k}_l$. The  masking process for each sequence  is expressed as $\bm{v}_l^o=\bm{r}_l^o \odot \bm{m}_l^o$, where $\bm{m}_l^o=[1,1,\ldots,1,0,\ldots,0]$ is the mask vector, with the first $k_l^o$ elements set to $1$ and the remaining elements set to $0$. Here,  $\odot$ denotes  element-wise multiplication. Finally, the output features are regrouped into  $M_l$ fixed-size blocks $\{{\bm{v}_{l,m}\}^{M_l}_{m=1}}$, with zero padding applied at the end if necessary. These blocks are then fed into the fine-grained encoder for further processing.

    \subsection{Fine-Grained Coding Design}
	To capture real-time channel variations, we  design the fine-grained encoder and decoder that utilize instantaneous SNR to dynamically refine the encoding strategy. As illustrated in Fig. \ref{fig:Fine_Encoder}, the fine-grained encoder  and decoder primarily  include the attention feature power allocation (AFPA) module, the unsqueeze and squeeze modules, and the attention feature re-encoding (AFR) module. A detailed explanation of these modules is provided below. We mainly focus on the encoder, the input of which is the remaining untransmitted
	blocks with a size of  $(N_l \times N_s \times L_e)$, where $N_l$ represents the number of remaining untransmitted blocks, $N_s$ denotes the number of OFDM symbols in each block, and $L_e=2L_f$ represents twice the number of subcarriers,  as each complex-valued channel symbol  is represented by two real-valued symbols.
	
	\textit{1) AFPA Module:} 
	The detailed architecture of the AFPA module is presented in the upper right of Fig. \ref{fig:Fine_Encoder}. This module is designed  to perform power allocation, inspired by the observation that the mean SNR of each subcarrier, i.e., $\mathbb{E}[\gamma^k_{l,m}]= \mathbb{E}[20\log(h^k_{l,m}\sqrt{P}/\sigma )]$, varies.
	Leveraging this statistical characteristic, the AFPA module assigns different weights to subcarriers, enabling adaptive power allocation to improve system performance.
	For example, when instantaneous channel conditions are poor, power can be  concentrated on high-SNR subcarriers to transmit critical information. To implement this, the AFPA module first transforms the instantaneous SNR into a latent representation via a sinusoidal embedding operation. It then processes this representation to generate scaled  and shifted vectors for each subcarrier. By applying these scaled and shifted vectors to the input features, power is adaptively allocated to each subcarrier based on the instantaneous SNR.
	
	\textit{2) Unsqueeze and Squeeze Modules:}
	Furthermore, considering that the channel conditions significantly differ across various blocks, we aim to integrate information from all blocks to enable the network to capture more complex features, thereby improving its robustness against varying channel conditions. To achieve this, we design the unsqueeze and squeeze modules, as illustrated in the upper right of Fig. \ref{fig:Fine_Encoder}. Specifically,
    in the unsqueeze module, we first concatenate the blocks to form  features of size
	$(N_e \times L_e)$, where $N_e=N_l \times N_s$. Then, the features undergo an unsqueeze operation, transforming the size to $(1 \times N_e \times L_e)$. The output of the unsqueeze module  is subsequently processed by two-dimensional convolution layers for feature extraction. The squeeze module  reverses this operation, i.e., transforming the processed features back to the original size of $(N_l \times N_s \times L_e)$.
	
	\textit{3) AFR Module:}
	We further propose an AFR module to embed  instantaneous SNR for   channel adaptation, as illustrated in the lower right of  Fig. \ref{fig:Fine_Encoder}.  The AFR module  takes the instantaneous SNR and  features of size  $(C_e \times N_e \times L_e)$ as input.
	Unlike the previous AFE  and AFPA  modules, the AFR module simultaneously accounts for the  statistical characteristics of subcarriers and the varying importance of features.
	This approach is particularly effective when there is a significant disparity between the instantaneous SNR and the average SNR, allowing for a more precise refinement of the encoding strategy. Additionally,  the AFR module incorporates a residual structure \cite{ResNet},  enabling  the model to retain existing encoding strategies when SNR variations are minimal, thereby mitigating potential losses from excessive fine-tuning.

	\subsection{Training Algorithm}
	We  present the detailed training algorithm  as follows.  Each data sample  
	$\bm{x}$ is  drawn  from a given dataset, with  its corresponding average SNR $\gamma_\textrm{{avg}}$  and  channel gain response  $\bm{H}_{l,m}$ of the $m$-th  block at the $l$-th layer sampled from predefined distributions. 
	Given an input image $\bm{x}$, we first sample an average SNR from the predefined distribution.
	Then, we process $\bm{x}$ through the bottom-up path and the top-down path to generate a set of latent representations $\bm{\mu} =\left\{\bm{\mu}_1, \bm{\mu}_2, \ldots, \bm{\mu}_L\right\}$.  For each  latent  representation, 
	we perform  coarse-grained  and  fine-grained encoding to generate  channel symbol vectors,  with the instantaneous channel gain response also sampled from the predefined distribution.  These channel symbol vectors are  transmitted  through wireless channels to the receiver. At the receiver, we perform fine-grained  and coarse-grained decoding of the  received channel symbol vectors.
In this way, we obtain all the reconstructed  representations $\bm{\hat{\mu}} =\left\{\bm{\hat{\mu}}_1, \bm{\hat{\mu}}_2, \ldots, \bm{\hat{\mu}}_L\right\}$.  Finally, by utilizing the top-down path at the receiver, the reconstructed image $\bm{\hat{x}}_H$ is   generated.

To jointly train the encoder and decoder via back-propagation, we employ the loss function as follows \cite{HVAE}:
\begin{equation}
	\mathcal{L}=
	\mathbb{E}_{
	\bm{x}\sim p_{\bm{x}},\bm{z}\sim q_{\bm{z} |\bm{x}}}
   \left[  \sum_{l=1}^{L} D_{K L}\left(q_{\bm{z}_l | \bm{z}_{<l}  ,\bm{x}} \| p_{\bm{z}_l | \bm{z}_{<l}}\right)\!-\!\log  p_{\bm{x} | \bm{z}} \right].
\end{equation}
We set the posterior $q_{\bm{z}_l | \bm{z}_{<l} ,\bm{x}}$   to a uniform distribution:
\begin{equation}
	q_{\bm{z}_l | \bm{z}_{<l} ,\bm{x}}(\bm{z}_l | \bm{z}_{<l} ,\bm{x}) \triangleq \prod_i\mathcal{U}\left(\mu_l^{(i)}-\tfrac{1}{2}, \mu_l^{(i)}+\tfrac{1}{2}\right),
\end{equation}
and  the prior $p_{\bm{z}_l | \bm{z}_{<l}} $  to a Gaussian distribution convolved with a uniform distribution \cite{hyperprior}:
\begin{equation}
	p_{\bm{z}_l | \bm{z}_{<l}} (\bm{z}_l | \bm{z}_{<l}) \triangleq \prod_i\mathcal{N}\left(\tilde{\mu}_l^{(i)}, (\tilde{\sigma}_l^{(i)})^2\right) * \mathcal{U}\left(-\tfrac{1}{2}, \tfrac{1}{2}\right),
\end{equation}
where $\mu_l^{(i)}$, $\tilde{\mu}_l^{(i)}$, and $\tilde{\sigma}_l^{(i)}$ denote the $i$-th element of $\bm{\mu}_l$, $\bm{\tilde{\mu}}_l$, and $\bm{\tilde{\sigma}}_l$, respectively. Relevant variables and calculation methods can be found in Fig. \ref{fig:Overall_architecture}. In this way, the loss function can be further simplified into
\begin{equation}
	\begin{aligned}
		\mathcal{L} 
		& =\mathbb{E}_{\bm{x}\sim p_{\bm{x}},\bm{z}\sim q_{\bm{z} |\bm{x}} }\left[\sum_{l=1}^L -\log p_{\bm{z}_l | \bm{z}_{<l}}(\bm{z}_l | \bm{z}_{<l})+\lambda \cdot d(\bm{x}, \bm{\hat{x}})\right],
	\end{aligned}
\end{equation} 
where $-\log {p_{\bm{z}_l | \bm{z}_{<l}}(\bm{z}_l |  \bm{z}_{<l})}$ represents the transmission rate,  $d(\bm{x}, \bm{\hat{x}})$ denotes the MSE distortion between  $\bm{x}$ and $\hat{\bm{x}}$, and $\lambda$ is the introduced weight to control the tradeoff between rate and distortion.
Subsequently, inspired by \cite{HJSCC}, we further extend the loss function to
\begin{equation}  \label{loss_function} 
	\begin{aligned}
		\mathcal{L}	
		&=  \mathbb{E}_{\bm{x}\sim p_{\bm{x}},\bm{z}\sim q_{\bm{z} |\bm{x}} }\big[ \\
		&\quad \sum_{l=1}^L - \alpha \log p_{\bm{z}_l | \bm{z}_{<l}}(\bm{z}_l | \bm{z}_{<l})+\lambda \cdot (d(\bm{x}, \bm{\hat{x}}) +  d(\bm{x}, \bm{\hat{x}}_{\text{H}})) \big],
	\end{aligned}
\end{equation}
where the additional distortion term $d(\bm{x}, \bm{\hat{x}}_{\text{H}})$
is introduced to optimize the transmission distortion, and the scaling parameter $\alpha$ is used to control the relationship between the entropy of the latent variable $\bm{z}_l$ and the transmission rate for $\bm{s}_l$. Moreover, the detailed training procedure is summarized in Algorithm 1.

\begin{algorithm}
	\DontPrintSemicolon
	\SetAlgoLined
	\KwIn {The training dataset $S$, epochs $Q$, the number of layers in CFA-JSCC $L$.}
	\KwOut {The parameters of the trained model.}
	\For{$q \gets 1$ \KwTo $Q$}{
		Sample a batch of data from $S$.\;
		Sample an average SNR.\;
		Compute latent representations  through the bottom-up path and top-down path.\;
		\For{$l \gets 1$ \KwTo $L$}{
			Compute  coarse-grained encoded symbols based on (\ref{coarse_encode}).\;
			Split  the output symbols into $M_l$ blocks.\;
			\For{$m \gets 1$ \KwTo $M_l$}{
				Sample channel gain response $\bm{H}_{l,m}$.\; 
				Compute instantaneous SNR based on (\ref{inst_snr}).\;
				Compute fine-grained encoded symbols based on (\ref{fine_encoder}).\;
				Compute received symbols based on (\ref{OFDM}).\;
				Compute fine-grained decoded symbols based on (\ref{fine_decoder}).\;
			}
			Integrate all the blocks.\;
			Compute coarse-grained decoded symbols based on (\ref{coarse_decoder}).\;
			Reconstruct the image using the top-down path.\;
			Compute the loss based on (\ref{loss_function}).\;
			Update the parameters of model.
		}
	}
	\caption{Training algorithm for the CFA-JSCC}
\end{algorithm}

\section{RL-Based CQI Selection}
	The CFA-JSCC relies on average SNR and instantaneous SNR  from the feedback link to adjust encoding strategies.
	If infinite-precision SNR is adopted, the feedback overhead would be unacceptable for practical communication systems.
	To address this issue, we consider quantizing the infinite-precision SNR. Specifically,
	for the average SNR, uniform quantization is adopted for simplicity, since system performance shows relative insensitivity to quantization loss in this case. In contrast, instantaneous SNR, which is required for each block, has a significant impact on model performance, making it necessary to design an efficient quantization
	method. To this end,
	we propose a novel RL-based CQI selection strategy,  where CQI is a discrete value drawn from the optional set  $\mathcal{Q}=\{0,1,\ldots,2^{B}-1\}$ with $B$ being the number of bits fed back to the transmitter.  Our proposed method primarily involves two mappings:
	\begin{itemize}
		\item[(i)]	
		\textit{SNR-to-CQI mapping:} At the receiver,   the continuous SNR $\gamma_{l,m}$  is converted into a discrete CQI. 
		\item[(ii)]
		\textit{CQI-to-SNR demapping:} At the transmitter,    the discrete CQI  is mapped back to a  quantized SNR $\hat{\gamma}_{l,m}$, which is subsequently fed into the fine-grained encoder.
	\end{itemize}
	\subsection{SNR-to-CQI Mapping}
	The SNR-to-CQI mapping transforms the infinite-precision SNR into a finite CQI, allowing only a small number of bits to be fed back for channel adaptation. This mapping is implemented using an RL-based method.
	
	\textit{1) Motivation for RL:} 
	When designing the SNR-to-CQI mapping, an intuitive approach is to treat it as a classification problem and solve it using a DL-based method, where the input is the SNR and the label is the CQI value. However, acquiring accurate labels is challenging. This is because  the SNR-to-CQI mapping  occurs for each individual block, while existing quality evaluation metrics, such as peak signal-to-noise ratio (PSNR), are designed for evaluating the quality of an entire image.  These metrics can only evaluate the  overall   effectiveness of multiple SNR-to-CQI mappings and cannot determine the optimal CQI for a specific SNR.
	
	The RL method with sparse rewards provides  a novel  perspective on SNR-to-CQI mapping. Similar to typical RL methods, it encompasses four key components: agents, states, actions, and rewards. However,  in the case of sparse rewards, the rewards are provided intermittently, i.e., only after achieving a specific objective.
	This  aligns naturally with the SNR-to-CQI mapping  task, as performance  metrics like PSNR (the reward) can only be obtained after the entire image has been transmitted. Therefore, we model the SNR-to-CQI mapping  as a sparse reward problem. Inspired by \cite{reward_shaping}, we propose a novel reward shaping scheme to address   this challenge, thereby facilitating the design of the SNR-to-CQI mapping.

	\textit{2) A Brief Overview of RL:}
	 Before  detailing  the proposed RL-based SNR-to-CQI mapping design, we commence with a brief overview of RL. RL  is an important branch  of machine learning, in which an agent continuously interacts with the environment to learn an optimal policy $\pi^*(\bm{t})$ that maximizes the expected discounted cumulative rewards, where $\bm{t}$ represents the current state.
	 In our design, we  focus on the model-free RL method Q-learning \cite{QL}, which introduces the action-value function $Q(\bm{t},\bm{a})$ to characterize the long-term cumulative reward obtained by taking action $\bm{a}$ in state $\bm{t}$. The core idea behind Q-learning is to iteratively update the Q-function based on the Bellman equation, which is given by
	 \begin{equation}
	 	Q(\bm{t},\bm{a}) \leftarrow
	 	Q(\bm{t},\bm{a}) 
	 	+ \rho \big[ \bm{r}+\kappa \max_{\bm{a}' \in \mathcal{A}} Q(\bm{t}',\bm{a}')-Q(\bm{t},\bm{a})\big],
	 \end{equation}
	 where $\rho \in (0,1)$ represents the learning rate, $\kappa \in (0,1)$ is the discount factor, $\bm{t}'$ denotes the next state, $\mathcal{A}$ is the set of possible actions, and  $\bm{r}$ is the reward for taking action $\bm{a}$ in state $\bm{t}$.
	 Although Q-learning has achieved notable success, it  struggles with continuous state spaces due to the infinite number of states.  To address this, the deep Q-network (DQN) \cite{DQN} is introduced, utilizing DNNs to approximate the Q-function. The Q-function is updated as 
	 \begin{equation} \label{Q_update}
	 	Q(\bm{t},\bm{a};\bm{\varphi})\leftarrow
	 	\bm{r}+\kappa \max_{\bm{a}' \in \mathcal{A}} Q(\bm{t}',\bm{a}';\bm{\varphi}),
	 \end{equation}
	 where $\bm{\varphi}$ are the trainable parameters of DQN, and $\bm{a}$ is given by the epsilon-greedy algorithm:
	 \begin{equation} \label{epsilon}
	 	\bm{a} =
	 	\begin{cases}
	 		\text{a random action in } \mathcal{A}, \text{   with prob. } \epsilon, \\
	 		
	 		\mathop{\arg\max}\limits_{\bm{a} \in \mathcal{A}} Q(\bm{t} ,\bm{a}; \bm{\varphi}), \text{ with prob. } 1 - \epsilon,
	 	\end{cases}
	 \end{equation}
	 where $\epsilon$  gradually decreases as the training procedure progresses.
	 
	 Furthermore, two key designs are introduced to improve the efficiency and stability of DQN training \cite{DQN}. First, DQN incorporates a target network to  support the training of the main network. Both networks share the same architecture, but the parameters of the  target network $\bm{\varphi}^-$ are copied from the main network at regular intervals and remain unchanged in between.
	 Secondly, to  further improve training efficiency, DQN incorporates experience replay. In this approach, the agent stores its interaction experiences ($\bm{t},\bm{a},\bm{r},\bm{t}'$) in an experience replay buffer. Instead of using only the most recent interaction  to update the network, DQN randomly samples a mini-batch of experiences from the  replay buffer for training.  Building on  these designs, the loss function is given by \cite{DQN}:
	 \begin{equation} \label{loss_DQN}
	 	\mathcal{L}_r=\frac{1}{D} \sum_{i=1}^{D} \big( \bm{r}_i+\kappa \max_{\bm{a}' \in \mathcal{A}} Q(\bm{t}'_i,\bm{a}'_i;\bm{\varphi}^-)-Q(\bm{t}_i,\bm{a}_i;\bm{\varphi}) \big)^2,
	 \end{equation}
	 where $D$ is the  mini-batch  size sampled  from the replay buffer.
	 
	\textit{3) RL-Based SNR-to-CQI Mapping Design:} 
	In this subsection, we present the RL-based  SNR-to-CQI mapping design. The key components of RL lie in the agent, action, state, and reward. In our design, focusing on the feedback process associated with the $m$-th block at the $l$-th layer, we define the agent as the decoder, the action as the CQI selection $\bm{a}_{l,m}$, and the state as the CSI $\boldsymbol{t}_{l,m}=\{\gamma_{\textrm{avg}},\gamma^k_{l,m}\}_{k=1}^{L_f}$, where $\gamma_{\textrm{avg}}$  is the average SNR, and $\gamma^k_{l,m}$ for $k= 1,2,\ldots,L_f$  represents the SNR of the  subcarrier. 
	
\begin{figure}[t]
		\centering 
		\includegraphics[width=0.98\linewidth]{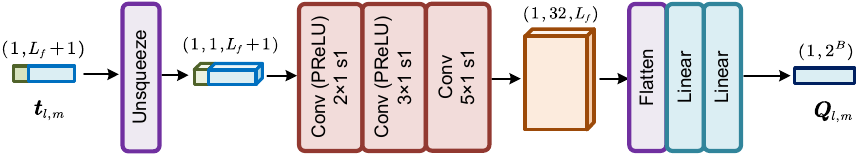} 
		\captionsetup{font={footnotesize  }}
		\captionsetup{justification=raggedright,singlelinecheck=false}
		\caption{ The network architecture of the DQN.}
		\label{fig:DQN} 
	\end{figure}

	The main challenge centers on the reward design.  In the RL method with sparse rewards, it is difficult to obtain immediate reward feedback for a given state-action pair. 
	To address  this challenge, we introduce a novel reward shaping scheme. Specifically, we  incorporate  two additional  rewards to guide  the agent's  learning: the PSNR and the SNR deviation. \text{PSNR} measures  the  overall   effectiveness of multiple SNR-to-CQI mappings, which is defined by
	\begin{equation} \label{PSNR}
		\text{PSNR}=10\log _{10}\frac{\text{MAX}^{2}}{\text{MSE}} (\text{dB}),
	\end{equation}
	where MSE $=d(\bm x,\bm {\hat{x}}_H)$ denotes the MSE between the source image $\bm x$ and the reconstructed image 
	$\bm{\hat{x}}_H$, and
	MAX is the maximum  pixel value. The SNR deviation  characterizes  the quantization loss  in a single SNR-to-CQI mapping, which  is expressed as
	\begin{equation}
		d_r(\gamma_{l,m},\hat{\gamma}_{l,m})=|\gamma_{l,m}-\hat{\gamma}_{l,m}| (\text{dB}),
	\end{equation}
	where $\gamma_{l,m}$ represents the SNR estimated by the receiver using the EESM algorithm, and $\hat{\gamma}_{l,m}$ denotes the quantized SNR at the transmitter. The reasons behind this design can be summarized as follows:
	\begin{itemize}
		\item[(i)]
		\textit{Relying on SNR deviation for preliminary learning:} During the initial training phase, the epsilon-greedy algorithm, as defined by Eq. (\ref{epsilon}), causes the SNR-to-CQI mapping  to be nearly random, resulting  in a generally low PSNR. This low PSNR fails to provide sufficient information to effectively guide the learning process. In contrast, the SNR deviation  characterizes  the quantization loss  in SNR, providing a more reliable metric that can effectively support the learning process during this phase.
		\item[(ii)]
		\textit{Relying on PSNR for refining  strategy:}
		In the later stages of training,  relying solely on the SNR deviation leads to a suboptimal SNR-to-CQI mapping strategy, as it does not account for the differing impacts of SNR mismatches at low and high SNR levels. In contrast,  since the model has already  learned  a preliminary SNR-to-CQI mapping,  the PSNR is no longer as low as before. It reflects the differences in CQI selection and provides valuable feedback to guide further learning. Therefore, we incorporate the PSNR to refine the mapping strategy and enhance system performance.
	\end{itemize}
	Building upon the  design  outlined above, the reward $\bm{r}_{l,m}$ for the $m$-th block at the $l$-th layer is defined as
	\begin{equation} \label{reward}
		\bm{r}_{l,m}=\xi(\text{PSNR} +\eta |\gamma_{l,m}-\hat{\gamma}_{l,m}|),
	\end{equation}
	where PSNR is shared  across all blocks within an image, $\xi$ represents the normalization parameter, and $\eta$ adjusts the weight between the PSNR and the SNR deviation,  which gradually decreases as the training procedure progresses. 
	 In this way, for an image with 
	$M=\sum_{l=1}^{L}M_l$ blocks, we can form $M$ experiences ($\bm{t}_{l,m},\bm{a}_{l,m},\bm{r}_{l,m},\bm{t}'_{l,m}$). Furthermore, we introduce
	DQN to manage the continuous state space, with its detailed architecture shown in Fig. \ref{fig:DQN}. Using these experiences and DQN, we can iteratively update the Q-function  according to Eq. (\ref{loss_DQN}) to identify an efficient SNR-to-CQI mapping.
	\subsection{CQI-to-SNR Demapping and Alternating Training }
	\textit{1) CQI-to-SNR Demapping Design:}
	The CQI-to-SNR demapping is designed to transform the discrete CQI into the quantized  SNR, enabling seamless integration of the CQI selection strategy into the CFA-JSCC framework without the need for extensive fine-tuning of the models. 
	This demapping is implemented using two linear layers with parameters
 	$\bm{\phi}$. 
 	 The training process is summarized as follows:
 	\begin{itemize}
 		\item[(i)]
		During the training process, we fix the network parameters of the CFA-JSCC model and only optimize the  parameters of the CQI-to-SNR demapping.
 		\item[(ii)]
 		We then encode the image  $\bm{x}$ in a manner similar to Algorithm 1, except that the instantaneous SNR is provided by the CQI-to-SNR demapping.
 		\item[(iii)]
 		 Using the loss function defined by Eq. (\ref{loss_function}), we utilize gradient descent to optimize the parameters of the CQI-to-SNR demapping.
 	\end{itemize}
	In this manner, we can iteratively  update the parameters $\bm{\phi}$ to optimize the CQI-to-SNR demapping and enhance its efficiency.

 	\textit{2) Training Algorithm for RL-Based CQI Selection:}
 	To train the SNR-to-CQI mapping and CQI-to-SNR demapping simultaneously, we  adopt an alternating training scheme. 
    In the first step,  we  fix the parameters of the CQI-to-SNR demapping and utilize the RL-based method to optimize the SNR-to-CQI mapping,  with the reward  defined by Eq. (\ref{reward}). In the second step, we fix the parameters of the SNR-to-CQI mapping and utilize gradient descent to  optimize the CQI-to-SNR demapping,  using the loss function  defined by Eq. (\ref{loss_function}). This alternating process continues until both networks converge. The detailed training procedure is outlined in Algorithm 2.
	
	\begin{algorithm} 
		\fontsize{9}{10.9}\selectfont 
		\DontPrintSemicolon
		\SetAlgoLined
		\KwIn {The training dataset, alternating training epochs $Q$, each with $Q_1$ SNR-to-CQI mapping training epochs and $Q_2$ CQI-to-SNR demapping training epochs. Mini-batch size $D$,  target network update interval $T$, the number of layers in CFA-JSCC $L$.}
		\KwOut {Optimized parameters $\{\bm{\varphi}, \bm{\varphi}^-, \bm{\phi}\}$.}
		
		Initialize  parameters and experience replay buffer $\mathcal{B}$.\;
		\For{$q \gets 1$ \KwTo $Q$}{
			Fix parameters of CQI-to-SNR demapping $\bm{\phi}$.\;
			\For{$q_1 \gets 1$ \KwTo $Q_1$}{
				Sample a batch of data and an average SNR $\gamma_{\text{avg}}$.\;
				Compute the quantized average SNR $\hat{\gamma}_{\text{avg}}$.\;
				\For{$l \gets 1$ \KwTo $L$}{
				Compute coarse-grained encoded symbols with $\hat{\gamma}_{\text{avg}}$ and split the output  into $M_l$ blocks.\;
				\For{$m \gets 1$ \KwTo $M_l$}{
					Sample $\bm{H}_{l,m}$, compute instantaneous SNR $\gamma_{l,m}$,  and obtain state $\boldsymbol{t}_{l,m}$.\;			Choose action $\bm{a}_{l,m}$ with (\ref{epsilon}).\;
					Compute quantized instantaneous SNR $\hat{\gamma}_{l,m}$.\;
					Perform fine-grained encoding on the $m$-th block with $\hat{\gamma}_{l,m}$ and transmit the output.\;
				}}
				Reconstruct the image and compute the PSNR.\;
				Compute reward $\bm{r}_{l,m}$ for all the blocks with (\ref{reward}).\;
				Form $\sum_{l=1}^{L}M_l$ experiences ($\bm{t}_{l,m},\bm{a}_{l,m},\bm{r}_{l,m},\bm{t}'_{l,m}$), store them into $\mathcal{B}$, and group them as a set.\;
				Sample $D$ sets of experiences from $\mathcal{B}$, compute the loss  based on (\ref{loss_DQN}), and update $\bm{\varphi}$.\;  
				Every $T$ steps, 	update  $\bm{\varphi}^-=\bm{\varphi}$.

			}
			Fix parameters of SNR-to-CQI mapping $\bm{\varphi},\bm{\varphi}^-$.\;
			\For{$q_2 \gets 1$ \KwTo $Q_2$}{
				Sample a batch of data and an average SNR $\gamma_{\text{avg}}$.\;
			Compute the quantized average SNR $\hat{\gamma}_{\text{avg}}$.\;
			\For{$l \gets 1$ \KwTo $L$}{
				Obtain $M_l$ blocks with $\hat{\gamma}_{\text{avg}}$ as before.\;
			\For{$m \gets 1$ \KwTo $M_l$}{
				Sample $\bm{H}_{l,m}$ and obtain state $\bm{t}_{l,m}$.\;
				Choose action $\bm{a}_{l,m}$ via (\ref{epsilon}) with $\epsilon=0$.\;
				Compute quantized instantaneous SNR $\hat{\gamma}_{l,m}$.\; 
				Perform fine-grained encoding  on the $m$-th block with  $\hat{\gamma}_{l,m}$ and transmit the output.\;}}
				Reconstruct the image.\;
				Compute the loss with (\ref{loss_function}) and update  $\bm{\phi}$.\;
			}
			}

		\caption{Training algorithm for RL-based CQI selection strategy}
	\end{algorithm}

	\section{Simulation Results}
	\subsection{Simulation Setup}
	To compare our CFA-JSCC model with existing DL-based JSCC models, we implement the proposed CFA-JSCC utilizing the deep learning platform Pytorch\footnote{Code and model weights will be available upon acceptance.}. The Adam optimizer is employed, with the learning rate of $1\times10^{-5}$. Moreover, we set the batch size to $8$, and conduct training for $200$ epochs. 
	Recognizing that large-sized images are more likely to experience multiple fading events,  we select $50,000$ images from the ImageNet dataset, cropping them to $256 \times 256$ for training. For testing, we employ the Kodak dataset, which comprises $24$ images with resolutions of $512 \times 768$ or $768 \times 512$. 
	To evaluate the performance of CFA-JSCC, we utilize the widely adopted pixel-wise metric PSNR and the multi-scale structural similarity index measure (MS-SSIM), where PSNR is defined by Eq. (\ref{PSNR}). 
	
	In the OFDM system setting, we set the number of OFDM symbols $N_s$ to $8$, pilot symbols $N_p$ to $1$, the number of subcarriers $L_f$ to $64$, and the cyclic  prefix length $L_{cp}$ to $16$. To model multipath fading, we use the clustered-delay-line-A (CDL-A) channel model \cite{CDL_A}, which is one of the channel models for fifth generation ($5$G) new radio (NR) systems standardized by the 3rd  Generation Partnership Project ($3$GPP).
		\begin{figure*}[t]
		\centering 
		\includegraphics[width=0.72\linewidth]{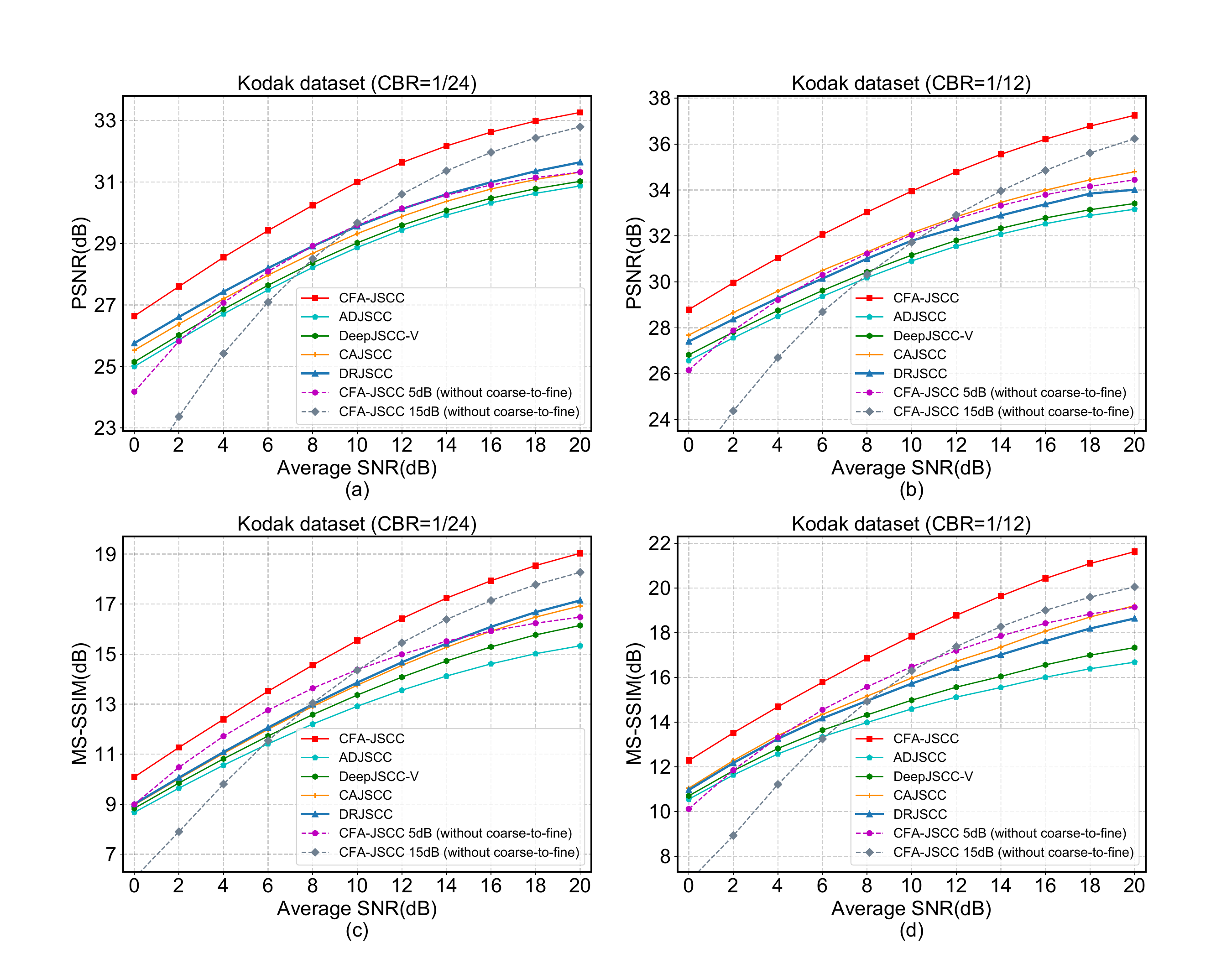} 
		\captionsetup{font={footnotesize  }}
		\captionsetup{justification=raggedright,singlelinecheck=false}
		\caption{The performance versus the average SNR under different CBRs and different metrics: (a) PSNR  with CBR set to $1/24$; (b) PSNR with CBR set to $1/12$; (c) MS-SSIM with CBR set to $1/24$; (d) MS-SSIM with CBR set to $1/12$.}
		\label{fig:task1_kodak} 
	\end{figure*}
	
	To verify the performance of our model, we compare CFA-JSCC with a series of models, including ADJSCC \cite{ADJSCC}, DeepJSCC-V \cite{DeepJSCC_V}, CAJSCC \cite{CAJSCC}, and DRJSCC \cite{DRJSCC}. To ensure a fair comparison, we extend all models to the OFDM system and adjust them to better  suit block fading channels. Specifically, for ADJSCC and DeepJSCC-V, which take a fixed SNR as input, we provide the models with the average SNR. For CAJSCC, which  requires both the average SNR and subcarrier channel gain coefficients  as inputs, we first calculate the average subcarrier channel gain coefficients across all fading events and then input them along with the average SNR. For DRJSCC, which considers multiple fading events, we input the instantaneous SNR, calculated  using the EESM algorithm, into the  model.  We train all  models  using average SNR values ranging from $0$ dB to $20$ dB. In addition, we  introduce  a variant of the CFA-JSCC model that  excludes the coarse-to-fine mechanisms. Specifically, we remove the AFE module from the coarse-grained encoder/decoder and the entire fine-grained encoder/decoder. This variant is trained with a fixed average SNR. Moreover, for CFA-JSCC and its variant, the set of permissible transmission lengths $\bm{k}_l$ needs to be transmitted to the receiver in an error-free manner. This overhead is relatively small compared to the overall overhead and  is included in the total overhead to ensure fairness.
	
			\begin{figure*}[t]
		\centering 
		\includegraphics[width=0.82\linewidth]{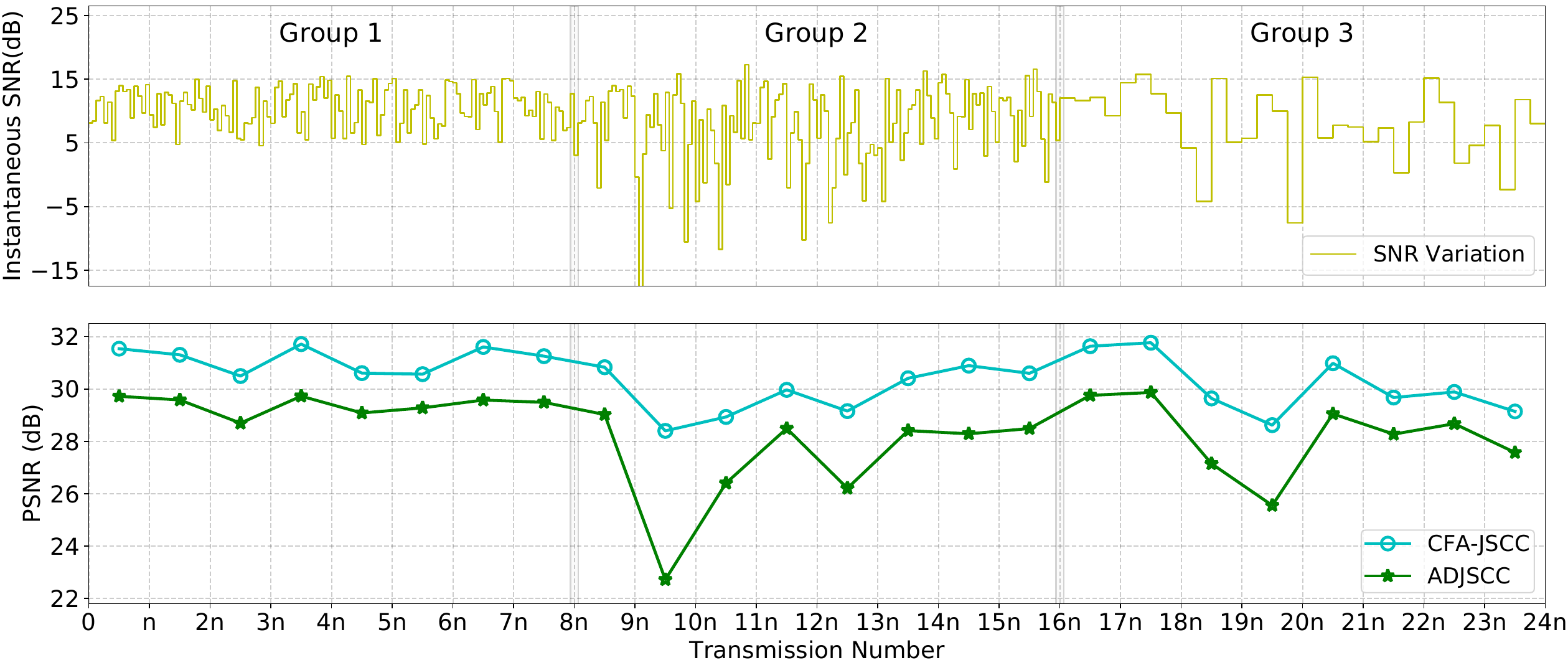} 
		\captionsetup{font={footnotesize  }}
		\captionsetup{justification=raggedright,singlelinecheck=false}
		\caption{ Visualization of the instantaneous SNR. The top part of the figure displays the variation of instantaneous SNR across $24$ constructed transmission cases and the bottom part of the figure presents the comparison results between CFA-JSCC and ADJSCC.}
		\label{fig:task2_test} 
	\end{figure*}
	
	\subsection{Performance Analysis}
	In Fig. \ref{fig:task1_kodak}, we display the transmission performance versus average SNR over 
	block fading channels, evaluated using PSNR and MS-SSIM.
	Specifically, Figs. \ref{fig:task1_kodak}(a) and   \ref{fig:task1_kodak}(b) illustrate the PSNR performance for CBR  values of $1/12$ and $1/24$, respectively.
	It is evident that our model outperforms the existing models that rely on fixed average SNR or channel gain coefficients as input, such as ADJSCC or CAJSCC. This demonstrates the advantage of utilizing instantaneous SNR for fine-grained channel adaptation, allowing our model to effectively  capture channel variations and adapt to time-varying channel conditions.
	Furthermore,   compared to DRJSCC, which considers multiple fading events during transmission, our model still achieves significantly superior performance. 
	This is because, when faced with significant channel fluctuations, the  refinements provided by DRJSCC  are not  flexible enough to   optimize the encoding strategy effectively.
	In contrast, CFA-JSCC employs a coarse-to-fine approach, integrating coarse-grained encoding for preliminary channel adaptation with fine-grained encoding for  dynamic refinement.
	This dual-phase approach demonstrates better robustness against significant channel fluctuations.
	Moreover,  compared to the variant of CFA-JSCC, our model  demonstrates  superior performance, even when the training  SNR  matches the  test SNR. This further highlights the advantages of the coarse-to-fine channel-adaptive strategy.
	Additionally, Figs. \ref{fig:task1_kodak}(c) and \ref{fig:task1_kodak}(d)  illustrate the MS-SSIM performance with respect to average SNR.  Since most MS-SSIM values exceed $0.9$, we convert them to dB for  improved clarity.  As with PSNR, our model significantly outperforms existing models. Furthermore,  when the average SNR increases, the performance gap between our model and existing channel-adaptive models becomes more pronounced. This improvement is attributed to the integration  of the hierarchical structure and the coarse-to-fine approach, enabling  our model to achieve better performance.

	To better illustrate the adaptability of our model to significant fluctuations and rapid changes in wireless channels, we visualize the variations in instantaneous SNR during the image transmission process.  Specifically,
	we  construct $24$ transmission cases in which the average SNR is fixed at $10$ dB and the instantaneous SNR adheres to the CDL-A model. These $24$ cases  are divided into three groups, each containing $8$ cases. In Group 1, the channel fluctuates rapidly with small amplitude; in Group 2, the channel fluctuates rapidly with large amplitude; and in Group 3,  the channel fluctuates slowly with large amplitude. Furthermore, a randomly selected image from the Kodak dataset is cropped to $256 \times 256$  pixels and then transmitted using both CFA-JSCC and ADJSCC in each of these cases.

	The top part of Fig. \ref{fig:task2_test} shows the variations in instantaneous SNR across the $24$ cases,  while the bottom  part of Fig. \ref{fig:task2_test} presents  a comparison between CFA-JSCC and ADJSCC. Each interval  $n$ represents a  different case.
	Comparing Group 1 with Group 2, we  observe that when the amplitude of channel fluctuations is relatively small, ADJSCC maintains moderate performance. However, as long as a block experiences significant channel fluctuations during transmission, the performance of ADJSCC  degrades substantially, even if  other blocks are only minimally affected by noise. In contrast, our model demonstrates strong adaptability to significant channel fluctuations, thanks to the preliminary channel adaptation provided by coarse-grained encoding, which establishes a solid foundation for handling such variations.
	In addition, comparing Group 2 with Group 3, we  observe that when the channel conditions change more rapidly, the performance gap between CFA-JSCC and ADJSCC becomes  more pronounced.
	This is because a fixed SNR can no longer effectively capture the variations in such dynamic channel scenarios.  As a result, the encoding strategy would fail to fit the channel environment.  In contrast, our model utilizes fine-grained encoding to dynamically refine  the encoding strategy,  ensuring it always aligns with channel conditions. Consequently, CFA-JSCC offers sufficient flexibility to handle rapid changes in wireless channels.

	\begin{figure}[t]
	\centering 
	\includegraphics[width=0.92\linewidth]{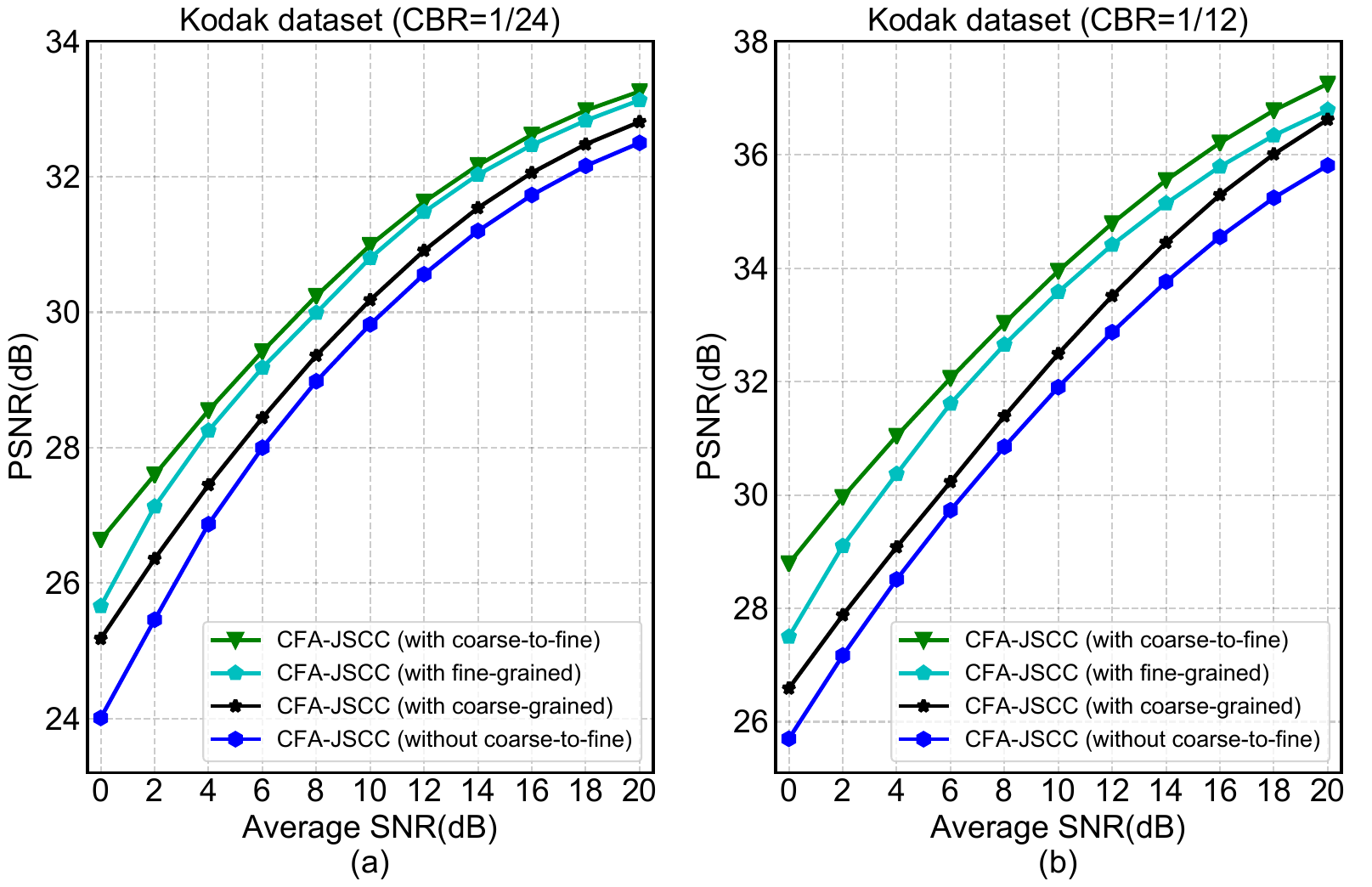} 
	\captionsetup{font={footnotesize  }}
	\captionsetup{justification=raggedright,singlelinecheck=false}
	\caption{Ablation studies on coarse-grained coding and fine-grained coding under the conditions of (a) $\text{CBR}=1/24$ and (b) $\text{CBR}=1/12$.}
	\label{fig:task3_ablation} 
\end{figure}

	\subsection{Ablation Studies}
	In this subsection, we demonstrate the effectiveness of the proposed coarse-grained coding  and fine-grained coding  methods through ablation studies.  Fig. \ref{fig:task3_ablation} presents the performance of CFA-JSCC under four different setups: 1) with coarse-to-fine coding, 2)  with fine-grained coding only, 3)  with coarse-grained coding only, and 4) without coarse-to-fine coding. 
	 All models are  trained with average SNR values ranging from $0$ dB to $20$ dB. 
	Figs. \ref{fig:task3_ablation}(a) and \ref{fig:task3_ablation}(b) show the results  for $\text{CBR}=1/12$ and $\text{CBR}=1/24$, respectively.
	It can be observed that, compared to the model without coarse-to-fine coding, the CFA-JSCC with the dual-phase channel-adaptive approach achieves significantly superior performance, with an average  improvement  of approximately $2$ dB. 
	Moreover, our model significantly outperforms the approach with coarse-grained coding only,   as the average SNR merely reflects the overall channel condition and fails to capture real-time channel variations effectively. 
	Furthermore, compared to the model with fine-grained coding only, our model also demonstrates  improved transmission performance, especially under low SNR conditions. This highlights the advantage of our coarse-to-fine method, which leverages a dual-phase approach to effectively handle significant channel fluctuations.

	\subsection{Results of RL-Based CQI Selection}
	
\begin{figure}[t]
		\centering 
		\includegraphics[width=0.98\linewidth]{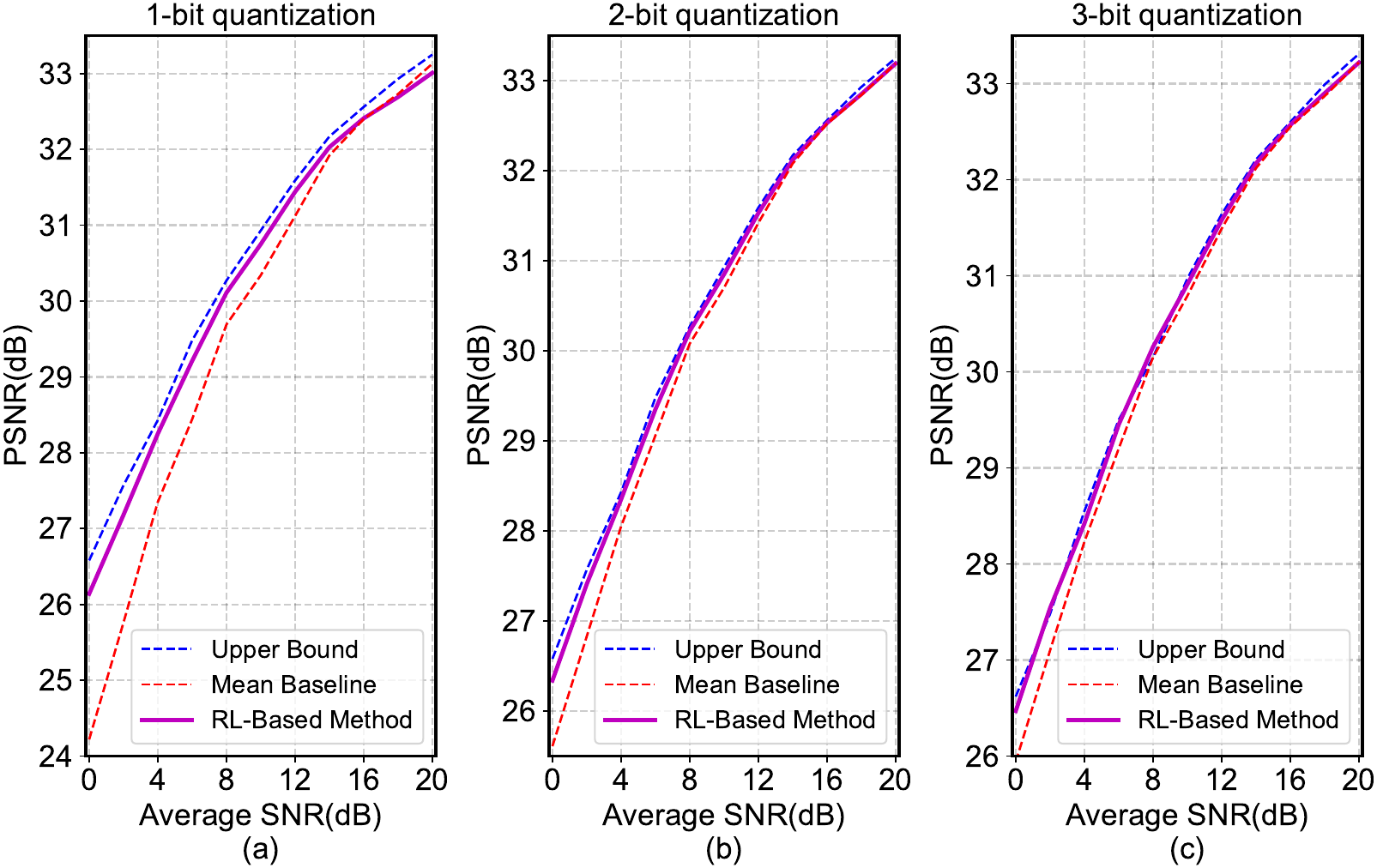} 
		\captionsetup{font={footnotesize  }}
		\captionsetup{justification=raggedright,singlelinecheck=false}
		\caption{Performance versus average SNR when using RL-based CQI selection strategy. The evaluation is based on (a) $1$-bit quantization, (b) $2$-bit quantization, and (c) $3$-bit quantization.}
		\label{fig:task4_RL} 
	\end{figure}

	In this subsection, we further evaluate the effectiveness of our RL-based CQI selection strategy.
	Figs. \ref{fig:task4_RL}(a), \ref{fig:task4_RL}(b), and  \ref{fig:task4_RL}(c) display the comparative results  for processing instantaneous SNR using $1$-bit, $2$-bit, and $3$-bit quantization, respectively.
	Specifically, for the  curve labeled ``Upper Bound", we use lossless average SNR and instantaneous SNR to transmit the image.
	For the curve labeled ``RL-Based Method", the average SNR is uniformly quantized into $3$ bits over the range of $0$ dB to $20$ dB, with the midpoint of each quantized interval  serving  as  input to the encoder. For example, all average SNR values from $0$ dB to $2.5$ dB  are mapped to $\text{CQI}=0$ at the receiver and then recovered to  $1.25$ dB at the transmitter for coarse-grained encoding.
	Regarding the instantaneous SNR in the RL scheme, it  is handled by our RL-based CQI selection strategy. 
	For the curve labeled ``Mean Baseline", the average SNR is processed in the same way as the ``RL-Based Method", but for the instantaneous SNR, we apply uniform quantization within the range of $-5$ dB to $25$ dB,  as most instantaneous SNR values fall within this range.
	
		\begin{figure*}[t]
		\centering 
		\includegraphics[width=0.81\linewidth]{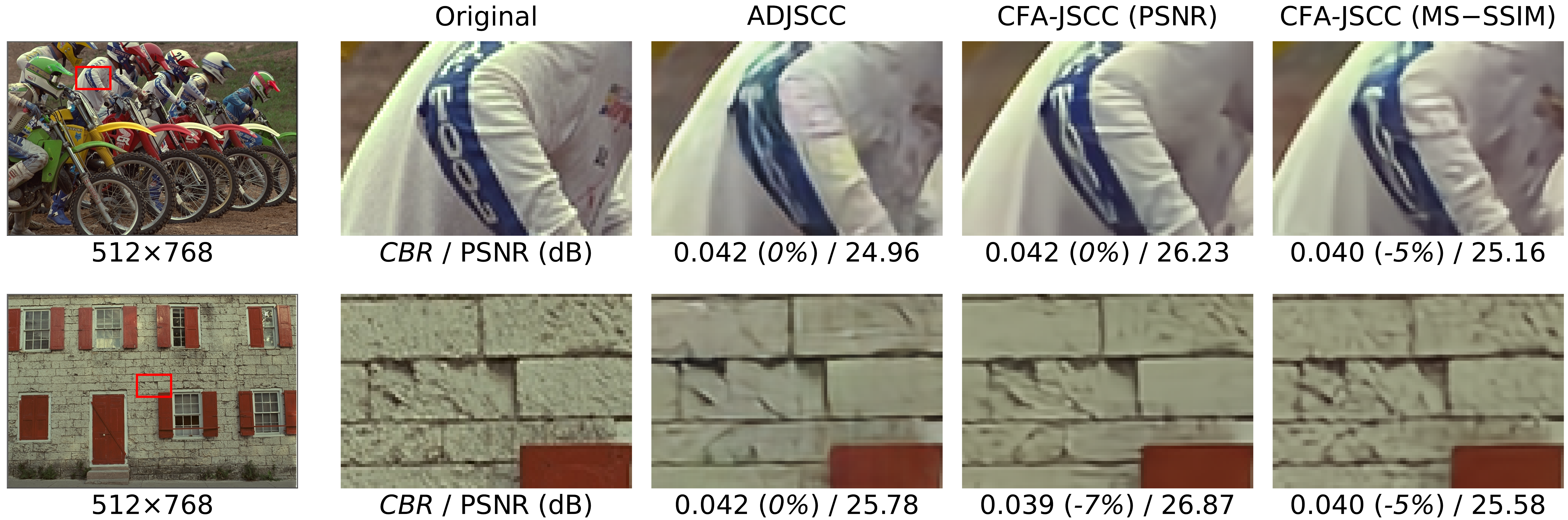} 
		\captionsetup{font={footnotesize  }}
		\captionsetup{justification=raggedright,singlelinecheck=false}
		\caption{Visual examples from different models. The first column presents the original images, the second column shows cropped patches of the original images, and the third to fifth columns display the reconstructed images generated by different models.}
		\label{fig:task6} 
	\end{figure*}

	From Fig. \ref{fig:task4_RL}(a), we observe that when only $1$-bit quantization is used for feedback, our proposed RL-based CQI selection strategy significantly outperforms the uniform quantization strategy, particularly at low average SNR. 
	This improvement  arises   from the ability of the RL-based CQI selection strategy  to optimize the quantization scheme. 
	Specifically, the strategy recognizes  that CFA-JSCC is more sensitive to SNR mismatch at low SNR and accordingly allocates finer quantization intervals in this regime,  thereby effectively  mitigating  the performance degradation caused by quantization.
	Furthermore, from Figs. \ref{fig:task4_RL}(b) and  \ref{fig:task4_RL}(c), we observe that as the number of quantization bits increases, the  performance  gap between our method and the  upper bound becomes  almost   negligible. However, the reduction in feedback overhead remains substantial, which aligns  with our goal of reducing  overhead while minimizing performance loss. 
	Additionally, we  find that our CFA-JSCC exhibits strong robustness against SNR  mismatches, especially at  higher  average SNR.
	
	\subsection{Visual Examples}
	To visually demonstrate the improvement in reconstruction quality provided by our model, we present a set of visual examples in Fig. \ref{fig:task6}. The first column presents the original images, the second column shows cropped patches of the original images, and the third to fifth columns display  the reconstructed images generated by different models. All reconstructed images are transmitted under  block fading channels with an average SNR of $10$ dB.
	 When comparing the third column (ADJSCC) with the fourth column (our model), we observe that our model offers better visual quality.
	Additionally, we compare models trained using PSNR and MS-SSIM. The model trained with the MS-SSIM metric excels at reconstructing fine details, such as folds in clothing or textures on walls, highlighting its superior visual quality.
	
	\section{Conclusion} \label{Conclusion}
	This paper considered wireless image transmission under time-varying block fading channels, where the transmission of a single image can experience multiple fading events. We proposed a novel CFA-JSCC framework to simultaneously  manage significant  fluctuations and rapid changes in wireless channels. Specifically, in the coarse-grained phase, CFA-JSCC adjusts the encoding strategy   for  preliminary channel adaptation. In the fine-grained phase, CFA-JSCC dynamically refines the encoding strategy to better align with evolving channel  conditions. Furthermore, to reduce feedback overhead, we developed an RL-based CQI selection strategy, where a reward shaping scheme is introduced to facilitate the training process. Simulation results demonstrated that our CFA-JSCC can offer enhanced flexibility  in capturing channel variations, leading to significant performance improvements.

	\bibliographystyle{IEEEtran}           
	\bibliography{IEEEabrv,Reference}      

\begin{thebibliography}{10}
\providecommand{\url}[1]{#1}
\csname url@samestyle\endcsname
\providecommand{\newblock}{\relax}
\providecommand{\bibinfo}[2]{#2}
\providecommand{\BIBentrySTDinterwordspacing}{\spaceskip=0pt\relax}
\providecommand{\BIBentryALTinterwordstretchfactor}{4}
\providecommand{\BIBentryALTinterwordspacing}{\spaceskip=\fontdimen2\font plus
\BIBentryALTinterwordstretchfactor\fontdimen3\font minus
  \fontdimen4\font\relax}
\providecommand{\BIBforeignlanguage}[2]{{%
\expandafter\ifx\csname l@#1\endcsname\relax
\typeout{** WARNING: IEEEtran.bst: No hyphenation pattern has been}%
\typeout{** loaded for the language `#1'. Using the pattern for}%
\typeout{** the default language instead.}%
\else
\language=\csname l@#1\endcsname
\fi
#2}}
\providecommand{\BIBdecl}{\relax}
\BIBdecl

\bibitem{DRJSCC}
J.~Pan, H.~Li, G.~Zhang, Y.~Cai, and G.~Yu, ``Deep refinement-based joint
  source channel coding over time-varying channels,'' in \emph{Proc. IEEE
  Wireless Commun. Netw. Conf. (WCNC)}, Apr. 2024, pp. 1--6.

\bibitem{iot}
M.~Mohammadi, A.~Al-Fuqaha, S.~Sorour, and M.~Guizani, ``Deep learning for
  {IoT} big data and streaming analytics: {A} survey,'' \emph{IEEE Commun.
  Surveys Tut.}, vol.~20, no.~4, pp. 2923--2960, 4th Quart. 2018.

\bibitem{data}
D.~Gündüz, Z.~Qin, I.~E. Aguerri, H.~S. Dhillon, Z.~Yang, A.~Yener, K.~K.
  Wong, and C.-B. Chae, ``Beyond transmitting bits: Context, semantics, and
  task-oriented communications,'' \emph{IEEE J. Select. Areas Commun.},
  vol.~41, no.~1, pp. 5--41, Jan. 2023.

\bibitem{TWC_1}
T.~Wu, Z.~Chen, D.~He, L.~Qian, Y.~Xu, M.~Tao, and W.~Zhang, ``{CDDM}: Channel
  denoising diffusion models for wireless semantic communications,'' \emph{IEEE
  Trans. Wireless Commun.}, vol.~23, no.~9, pp. 11\,168--11\,183, Sep. 2024.

\bibitem{TWC_2}
J.~Mao, K.~Xiong, M.~Liu, Z.~Qin, W.~Chen, P.~Fan, and K.~B. Letaief, ``A
  {GAN}-based semantic communication for text without {CSI},'' \emph{IEEE
  Trans. Wireless Commun.}, vol.~23, no.~10, pp. 14\,498--14\,514, Oct. 2024.

\bibitem{TWC_3}
P.~Yi, Y.~Cao, X.~Kang, and Y.-C. Liang, ``Deep learning-empowered semantic
  communication systems with a shared knowledge base,'' \emph{IEEE Trans.
  Wireless Commun.}, vol.~23, no.~6, pp. 6174--6187, Jun. 2024.

\bibitem{DeepJSCC}
E.~Bourtsoulatze, D.~Burth~Kurka, and D.~Gündüz, ``{Deep joint source-channel
  coding for wireless image transmission},'' \emph{IEEE Trans. Cognit. Commun.
  Netw.}, vol.~5, no.~3, pp. 567--579, Sep. 2019.

\bibitem{Deep_F}
D.~B. Kurka and D.~Gündüz, ``{DeepJSCC}-f: {Deep} joint source-channel coding
  of images with feedback,'' \emph{IEEE J. Select. Areas Inf. Theory}, vol.~1,
  no.~1, pp. 178--193, May 2020.

\bibitem{perc}
J.~Wang, S.~Wang, J.~Dai, Z.~Si, D.~Zhou, and K.~Niu, ``Perceptual learned
  source-channel coding for high-fidelity image semantic transmission,'' in
  \emph{Proc. IEEE Global Commun. Conf. (GLOBECOM)}, Dec. 2022, pp. 3959--3964.

\bibitem{BI}
L.~Sun, Y.~Yang, M.~Chen, C.~Guo, W.~Saad, and H.~V. Poor, ``Adaptive
  information bottleneck guided joint source and channel coding for image
  transmission,'' \emph{IEEE J. Select. Areas Commun.}, vol.~41, no.~8, pp.
  2628--2644, Aug. 2023.

\bibitem{domain}
A.~Li, X.~Liu, G.~Wang, and P.~Zhang, ``Domain knowledge driven semantic
  communication for image transmission over wireless channels,'' \emph{IEEE
  Wireless Commun. Lett.}, vol.~12, no.~1, pp. 55--59, Jan. 2023.

\bibitem{JSCC_text}
H.~Xie, Z.~Qin, G.~Y. Li, and B.-H. Juang, ``Deep learning enabled semantic
  communication systems,'' \emph{IEEE Trans. Signal Process.}, vol.~69, pp.
  2663--2675, Apr. 2021.

\bibitem{Deep_video}
T.-Y. Tung and D.~Gündüz, ``{DeepWiVe: Deep-learning-aided wireless video
  transmission},'' \emph{IEEE J. Sel. Areas Commun.}, vol.~40, no.~9, pp.
  2570--2583, Sep. 2022.

\bibitem{Deep_mul}
G.~Zhang, Q.~Hu, Z.~Qin, Y.~Cai, G.~Yu, and X.~Tao, ``A unified multi-task
  semantic communication system for multimodal data,'' \emph{IEEE Trans.
  Commun.}, vol.~72, no.~7, pp. 4101--4116, Jul. 2024.

\bibitem{JSCC_speech}
Z.~Weng and Z.~Qin, ``Semantic communication systems for speech transmission,''
  \emph{IEEE J. Select. Areas Commun.}, vol.~39, no.~8, pp. 2434--2444, Aug.
  2021.

\bibitem{JSCC_perc}
J.~Wang, S.~Wang, J.~Dai, Z.~Si, D.~Zhou, and K.~Niu, ``Perceptual learned
  source-channel coding for high-fidelity image semantic transmission,'' in
  \emph{Proc. IEEE Global Commun. Conf. (GLOBECOM)}, Dec. 2022, pp. 3959--3964.

\bibitem{TASK_O}
H.~Xie, Z.~Qin, X.~Tao, and K.~B. Letaief, ``Task-oriented multi-user semantic
  communications,'' \emph{IEEE J. Select. Areas Commun.}, vol.~40, no.~9, pp.
  2584--2597, Sep. 2022.

\bibitem{DL_perc}
D.~Huang, X.~Tao, F.~Gao, and J.~Lu, ``Deep learning-based image semantic
  coding for semantic communications,'' in \emph{Proc. IEEE Global Commun.
  Conf. (GLOBECOM)}, Dec. 2021, pp. 1--6.

\bibitem{HARQ}
P.~Jiang, C.-K. Wen, S.~Jin, and G.~Y. Li, ``Deep source-channel coding for
  sentence semantic transmission with {HARQ},'' \emph{IEEE Trans. Commun.},
  vol.~70, no.~8, pp. 5225--5240, Aug. 2022.

\bibitem{ADJSCC}
J.~Xu, B.~Ai, W.~Chen, A.~Yang, P.~Sun, and M.~Rodrigues, ``{Wireless image
  transmission using deep source channel coding with attention modules},''
  \emph{IEEE Trans. Circuits Syst. Video Technol.}, vol.~32, no.~4, pp.
  2315--2328, Apr. 2022.

\bibitem{DeepJSCC_V}
W.~Zhang, H.~Zhang, H.~Ma, H.~Shao, N.~Wang, and V.~C.~M. Leung, ``Predictive
  and adaptive deep coding for wireless image transmission in semantic
  communication,'' \emph{IEEE Trans. Wireless Commun.}, vol.~22, no.~8, pp.
  5486--5501, Aug. 2023.

\bibitem{SCAN}
G.~Zhang, Q.~Hu, Y.~Cai, and G.~Yu, ``{SCAN}: Semantic communication with
  adaptive channel feedback,'' \emph{IEEE Trans. Cognit. Commun. Netw.},
  vol.~10, no.~5, pp. 1759--1773, Oct. 2024.

\bibitem{OFDM}
M.~Yang, C.~Bian, and H.-S. Kim, ``{OFDM}-guided deep joint source channel
  coding for wireless multipath fading channels,'' \emph{IEEE Trans. Cognit.
  Commun. Netw.}, vol.~8, no.~2, pp. 584--599, Jun. 2022.

\bibitem{CAJSCC}
H.~Wu, Y.~Shao, K.~Mikolajczyk, and D.~Gündüz, ``{Channel-adaptive wireless
  image transmission with OFDM},'' \emph{IEEE Wireless Commun. Lett.}, vol.~11,
  no.~11, pp. 2400--2404, Nov. 2022.

\bibitem{MITT}
K.~Yang, S.~Wang, J.~Dai, K.~Tan, K.~Niu, and P.~Zhang, ``{WITT}: A wireless
  image transmission transformer for semantic communications,'' in \emph{IEEE
  Int. Conf. Acoust. Speech and Signal Process. (ICASSP)}, Jun. 2023, pp. 1--5.

\bibitem{DL_MMSE}
C.~Karamanli, T.-Y. Tung, and D.~Gündüz, ``Model-driven deep joint
  source-channel coding over time-varying channels,'' in \emph{Proc. Int. ITG
  Workshop Smart Antennas (WSA) and Conf. Syst. Commun. Coding (SCC)}, Feb.
  2023, pp. 1--6.

\bibitem{DeepJSCC_l}
C.~Bian, Y.~Shao, and D.~Gündüz, ``{DeepJSCC}-l++: Robust and
  bandwidth-adaptive wireless image transmission,'' in \emph{Proc. IEEE Global
  Commun. Conf. (GLOBECOM)}, Dec. 2023, pp. 3148--3154.

\bibitem{ADJSCC_l}
X.~Bao, M.~Jiang, and H.~Zhang, ``{ADJSCC-l}: {SNR}-adaptive {JSCC} networks
  for multi-layer wireless image transmission,'' in \emph{Int. Conf. Comput.
  Commun. (ICCC)}, Dec. 2021, pp. 1812--1816.

\bibitem{swin_taming}
K.~Yang, S.~Wang, J.~Dai, X.~Qin, K.~Niu, and P.~Zhang, ``{SwinJSCC}: Taming
  swin transformer for deep joint source-channel coding,'' \emph{IEEE Trans.
  Cognit. Commun. Netw.}, pp. 1--1, to appear, 2024, doi:
  10.1109/TCCN.2024.3424842.

\bibitem{HJSCC}
G.~Zhang, H.~Li, Y.~Cai, G.~Yu, and R.~Zhang, ``Learned image transmission with
  hierarchical variational autoencoder,'' \emph{arXiv preprint
  arXiv:2408.16360}, 2024.

\bibitem{HVAE}
A.~Vahdat and J.~Kautz, ``{NVAE}: A deep hierarchical variational
  autoencoder,'' in \emph{Adv. Neural Inf. Process. Syst. (NeurIPS)}, vol.~33,
  Dec. 2020, pp. 19\,667--19\,679.

\bibitem{EESM}
R.~Sandanalakshmi, T.~Palanivelu, and K.~Manivannan, ``Effective {SNR} mapping
  for link error prediction in {OFDM} based systems,'' in \emph{Proc. IET-UK
  Int. Conf. lnf. Commun. Technol. Elect. Sci. (ICTES)}, Dec. 2007, pp.
  684--687.

\bibitem{conv_net}
Z.~Liu, H.~Mao, C.-Y. Wu, C.~Feichtenhofer, T.~Darrell, and S.~Xie, ``A
  {ConvNet} for the 2020s,'' in \emph{Proc. IEEE/CVF Conf. Comput. Vis. Pattern
  Recognit. (CVPR)}, Jun. 2022, pp. 11\,976--11\,986.

\bibitem{NTSCC}
J.~Dai, S.~Wang, K.~Tan, Z.~Si, X.~Qin, K.~Niu, and P.~Zhang, ``Nonlinear
  transform source-channel coding for semantic communications,'' \emph{IEEE J.
  Select. Areas Commun.}, vol.~40, no.~8, pp. 2300--2316, Aug. 2022.

\bibitem{ResNet}
K.~He, X.~Zhang, S.~Ren, and J.~Sun, ``{Deep residual learning for image
  recognition},'' in \emph{Proc. IEEE Conf. Comput. Vis. Pattern Recognit.
  (CVPR)}, Jun. 2016, pp. 770--778.

\bibitem{swin}
Z.~Liu, Y.~Lin, Y.~Cao, H.~Hu, Y.~Wei, Z.~Zhang, S.~Lin, and B.~Guo, ``Swin
  transformer: Hierarchical vision transformer using shifted windows,'' in
  \emph{Proc. IEEE/CVF Int. Conf. Comput. Vis. (ICCV)}, Oct. 2021, pp.
  10\,012--10\,022.

\bibitem{hyperprior}
J.~Ball{\'e}, D.~Minnen, S.~Singh, S.~J. Hwang, and N.~Johnston, ``Variational
  image compression with a scale hyperprior,'' \emph{arXiv preprint
  arXiv:1802.01436}, 2018.

\bibitem{reward_shaping}
A.~Y. Ng, D.~Harada, and S.~Russell, ``Policy invariance under reward
  transformations: Theory and application to reward shaping,'' in \emph{Proc.
  Int. Conf. Mach. Learn. (ICML)}, vol.~99, Jul. 1999, pp. 278--287.

\bibitem{QL}
C.~J. Watkins and P.~Dayan, ``Q-learning,'' \emph{Mach. learn.}, vol.~8, no.
  3-4, pp. 279--292, May 1992.

\bibitem{DQN}
V.~Mnih, K.~Kavukcuoglu, D.~Silver \emph{et~al.}, ``Human-level control through
  deep reinforcement learning,'' \emph{Nature}, vol. 518, no. 7540, pp.
  529--533, Feb. 2015.

\bibitem{CDL_A}
``Study on channel model for frequencies from 0.5 to 100 {GHz} (release 16)
  v16.1.0,'' {3GPP TR 38.901}, Tech. Rep., Nov. 2020.

\end{thebibliography}

\end{document}